\documentstyle[12pt,aaspp4]{article}
\input epsf

\newcommand\bomega{\mbox{\boldmath $\omega$}}

\def\ep{\epsilon}

\def\ni{\noindent}

\def\bar{\overline}

\def\emf{\overline{\mbox{${\cal E}$}} {}}

\def\emfb{\overline{\mbox{\boldmath ${\cal E}$}} {}}

\def\bfepo{\overline{\mbox{\boldmath ${\epsilon}$}} {}}

\def\epo{\overline{\mbox{${\epsilon}$}} {}}

\def\bfep{{\mbox{\boldmath ${\epsilon}$}} {}}

\def\bbE{\bar {\bf E}}

\def\beq{\begin{equation}}
\def\ee{\end{equation}}
\def\lsim{\mathrel{\rlap{\lower4pt\hbox{\hskip1pt$\sim$}}
    \raise1pt\hbox{$<$}}}
\def\gsim{\mathrel{\rlap{\lower4pt\hbox{\hskip1pt$\sim$}}
    \raise1pt\hbox{$>$}}}
\def\bfE{{\bf E}}
\def\bfJ{{\bf J}}

\def\bfA{{\bf A}}
\def\bfa{{\bf a}}
\def\bfe{{\bf e}}
\def\bfB{{\bf B}}
\def\bbJ{\bar {\bf J}}

\def\ts{\times}
\def\lb{\langle}
\def\rb{\rangle}
\def\curl{\nabla {\ts}}

\def\bbV{\bar {\bf V}}

\def\bfv{{\bf v}}
\def\bfvp{{\bf v}}

\def\bfj{{\bf j}}

\def\bfe{{\bf e}}
\def\bfw{{\bomega}}

\def\bfb{{\bf b}}

\def\bfB{{\bf B}}

\def\bbB{\overline {\bf B}}
\def\bbA{\overline {\bf A}}

\begin{document}

\setcounter{equation}{0}

\title{Bihelical magnetic relaxation and large scale magnetic field growth}


\author{Eric G. Blackman} 
\affil{ Department of Physics \& Astronomy and Laboratory for
Laser Energetics, University of Rochester, Rochester NY 14627;
email:  blackman@pas.rochester.edu}

\centerline {(submitted to  Physics of Plasmas)} 

\begin{abstract}


A unified, three-scale system of equations accommodating 
nonlinear velocity driven helical dynamos, as well as  
time-dependent relaxation of  magnetically dominated unihelical
or bihelical systems is derived and solved herein.  
When opposite magnetic helicities of equal magnitude are 
injected on the intermediate and small scales, 
the large scale magnetic helicity grows kinematically (independent
of the magnetic Reynolds number) to 
equal that on the intermediate scale. 
For both free and driven relaxation
large scale fields are rapidly produced.
Subsequently, a dissipation-limited dynamo,
driven by growth of small scale kinetic helicity, 
further amplifies the large scale field.
The results are important for astrophysical coronae fed with 
bihelical structures by dynamos in their host rotators.
The large scale for the rotator corresponds to the intermediate
scale for the corona. 
That bihelical magnetic relaxation can produce global scale fields
may help to explain the formation of  
astrophysical coronal holes and magnetohydrodynamic outflows.

\end{abstract}


\centerline{PACS codes: 52.30.Cv, 95.30.Qd, 
52.65Kj,  96.60.Pb, 98.35.Eg, 98.62Mx}




\section*{I. Introduction}

\section*{A. Background}

Large scale helical dynamo  theory 
provides a framework for modeling
the in situ origin of large-scale magnetic field 
growth in planets, stars, and galaxies$^{\cite{moffatt}-\cite{zeldovich83}}$
and has also been invoked 
and to explain the sustenance of fields in  fusion devices.$^{\cite{strauss85}-\cite{jiprager02}}$
Much of the associated dynamo 
research can be divided into 
(1) that which parameterizes the nonlinearities
in an effort to practically  model 
the field evolution of specific objects or systems, and 
(2) that aimed at identifying the 
physical principles of the nonlinear dynamo 
by dynamically including the mutual backreaction 
between  the field
and velocities  for simplified systems.

Progress in understanding the 
nonlinear helical field growth and saturation 
has resulted from a combination of numerical
and analytic work that dynamically 
incorporates magnetic helicity evolution.$^{\cite{pfl}-\cite{b03}}$
Analytic models$^{\cite{fb02}-\cite{matthaeus04}}$ 
employing a discrete number of scales 
and including the backreaction of the field growth on the turbulent 
velocity have shown that for 
kinetic helicity driven dynamos, the build up of small
scale magnetic helicity ultimately quenches the large
scale field growth, in agreement  with numerical
simulations.$^{\cite{b2001}-\cite{bs2004}}$.
A two-scale nonlinear theory of large scale field growth  
from a dynamo driven by small scale magnetic helicity fluctuations
rather than kinetic helicity fluctuations 
has also been developed.$^{\cite{bf04}}$
In this case the kinetic helicity becomes the quenching agent rather than
the growth driver.  Like the velocity-driven 
helical dynamo, such magnetically driven helical dynamos thrive on 
a finite  turbulent electromotive force  in Ohm's law.

A magnetically driven helical dynamo can also be described
as ``dynamical magnetic relaxation.''
Magnetic relaxation describes  the process 
by which magnetic structures in magnetic pressure dominated environments 
evolve to their equilibrium states. The fully relaxed  state is the Taylor 
state,$^{\cite{taylor86}}$ 
determined by minimizing the magnetic energy subject
to the constraint that magnetic helicity is conserved.
The result is a force-free helical configuration
with the  field relaxing to the largest scale available,  
subject to boundary conditions.
However, Taylor's theory by itself is not a dynamical theory 
since it does not provide a time-dependent  description of how 
the large scale magnetic helicity evolves.
A time-dependent theory that also includes velocity and fluctuations is required.  

In fusion devices, where the plasma is  magnetically 
dominated, the word ``dynamo'' is typically used to describe
a magnetically dominated dynamo in which  magnetic flux is converted from
toroidal to poloidal (or vice versa), increase the scale
of the field, and sustain a strong magnetic flux against
microphysical dissipation.$^{\cite{strauss85}-\cite{jiprager02}}$
Detailed studies of these dynamos 
focus  on the instabilities that drive the fluctuations (tearing modes)
as well as the subsequent relaxation, conversion and sustenance of the magnetic flux. The full boundary value problem can be solved for 
specific configurations.  

Similar magnetically driven dynamos are likely an 
important part of the generation of 
large scale astrophysical fields that mediate coronal holes and 
magnetohydrodynamic (MHD) jets in astrophysics.
The analogy can be drawn between spheromak or reversed-field pinch 
formation$^{\cite{bellan00}}$
and coronal field relaxation: The laboratory configurations form 
because the injection of magnetic helicity into a system  takes
the system away from the relaxed state, but also 
drives tearing mode fluctuations that yield a finite turbulent electromotive 
force. This in turn drives the system to  to a new  relaxed state.
In astrophysical coronae the situation
is similar. The helical field 
produced by a velocity
driven dynamo inside of a rotator is buoyantly fed  from below 
into the corona.
This takes the magnetically dominated corona away from the relaxed state 
initially, but fluctuations arise in which the system is then driven back
to a relaxed state.  The corona therefore acts as the 
``laboratory'' plasma volume,
injected with helicity from below.  In the astrophysical case,
the injection occurs not at one site, but throughout the 
coronal-surface boundary. 
While each of the individual injection
sites (flux tubes)  would be analogous to a single laboratory configuration
if the twists are injected along the tube, each tube can also
open up to even larger scales by mutual interaction. In this way,
magnetic energy can appear on scales even larger than those on
which the field was originally present. This is a difference
between the laboratory and astrophysical cases.

\section*{B. Astrophysical motivation for bihelical relaxation}

Magnetically driven dynamos
are important for a  wide range of 
astrophysical coronae because  all coronae 
of  turbulent astrophysical rotators 
(the sun, stars, and accretion disks) are likely magnetically dominated.
For the solar corona,$^{\cite{gp97}}$ observations support the 
basic paradigm that particle acceleration and heating
result from extraction of coronal magnetic energy
via reconnection or wave dissipation.$^{\cite{withbroe77,sz}}$
The coronal magnetic fields 
come from magnetic loops produced within the sun that are 
injected into the corona.  Because the field polarity reverses
every 11 years, these loops could not have been merely 
the residue of the accreted interstellar field.
The sun also has much larger
scale (``open'') fields in coronal holes along which the solar
wind propagates.  These fields also reverse during the solar
cycle, again highlighting that they are not 
simply the result of flux freezing from the sun's formation.  
That the geometry and topology of 
smaller bipolar structures evolve to produce global
structures in the solar cycle can be modeled empirically.$^{\cite{wang03}}$

Accretion disk coronae likely involve similar processes. 
Non-thermal emission and flares from accreting  
stars or black holes are also naturally
explained by the emergence
of magnetic fields from the turbulent plasma below.$^{\cite{galeev79}-\cite{wang04}}$
By analogy to the sun, disk fields supplied to its corona
can open up. Because accretion disks 
are ``fast rotators'' (as are young X-ray luminous stars$^{\cite{fm99}}$) 
the open field lines are not only the track  
along which outflows propagate, but 
can act as a drive belt to extract the rotational energy in the disk
(or young star) supplied by accretion.
Conversion of the accretion energy flux into 
Poynting flux can drive the observed outflows.$^{\cite{bp82}-\cite{lovelace02}}$
Though particular MHD models differ as to how far above the
disk the outflows are carried as Poynting flux,$^{\cite{bp82,lovelace02}}$
they all involve large scale fields and collectively offer   
the most viable jet/outflow paradigm for a range of accretion engines.

MHD jet/outflow studies typically impose the global field 
anchored in the turbulent rotator as an initial condition, 
and then allow the system to evolve or consider the steady state.$^{\cite{bp82}-\cite{vrb04}}$.  Because field reversal 
cycles cannot be directly measured in accretion disks, 
it is not as straightforward as it is for the sun to
directly determine whether the large 
scale fields are accreted or produced in situ.
But difficulties in accreting flux in a thin turbulent
disk$^{\cite{lpp},\cite{blackmantan04}}$  and the fact that 
the solar corona produces large scale global fields
in situ provide reasons to believe that 
in situ production in disk coronae is ongoing.
The supply of magnetic energy to a corona is
directly supported by 3-D MHD simulations,$^{\cite{ms00}}$ but the 
self-consistent generation of large scale flux still needs further study.

In the standard use of helical dynamos to generate large scale astrophysical 
fields,  an initially weak field is assumed 
and the growth of the large scale field 
is studied for given a choice of velocity 
driving and global boundary 
conditions.$^{\cite{zeldovich83},\cite{campbell99},\cite{vrb04},\cite{pudritz81}-\cite{rrs99}}$ While this approach applies inside the rotator,
it hides an important aspect of the large
scale field generation in the corona.  
The large scale coronal fields might be more appropriately described
as the second phase of a two-phase process: (1) In the first phase, a 
velocity driven helical dynamo generates large scale
fields inside the rotator. The field produced then 
emerges into  the corona.  
(In order for buoyancy to beat shredding for a disk, 
a helical dynamo within the rotator seems to be required because 
non-helical turbulent amplification produces 
structures that are dissipated
before they can rise significantly$^{\cite{blackmantan04}}$.)
(2) Because the ``large'' scale field produced by the velocity driven
dynamo in the disk is of ``intermediate'' scale with respect to the
corona, it must incur further relaxation to open up to the global
scales required by holes and outflows--just as we see in the sun.  However,
this relaxation occurs in an already
magnetically dominated environment, so phase (2),  
just discussed, is indeed a magnetically dominated dynamo. 
Because magnetic helicity
conservation implies that 
any nonlinear dynamo in the rotator below produces
helical fields of opposite signs on different scales, 
$^{\cite{pfl}-\cite{bb03}}$ (see Fig. 1 of Ref. \cite{bb03} for an 
illustration) the question becomes:
Is a net global large scale field produced when
injected with bihelical fields that supply a zero net helicity?

\section*{C. Present aim: closed time-dependent theory, not detailed boundary value study}

Unlike the laboratory dynamo, the astrophysical coronal dynamo
involves multiple helicity injection sites (i.e. each buoyant twisted flux
tube) rather than a single injection (helicity injection
along the mean field.).
Nevertheless, in a real corona, like in a  laboratory configuration,
the global magnetic relaxation will result from current instabilities
instigated by the magnetic helicity injection.$^{\cite{bellan00,bbs02}}$
Indeed, a full treatment of coronal bihelical magnetic relaxation
will ultimately require boundary value studies.
However, much insight into the basic nonlinear effects of 
magnetic helicity transfer and velocity backreaction
can be gained using a minimalist mean field approach
 without a detailed modeling of the tearing modes 
that drive the fluctuations.  Rather than a detailed boundary
value problem, I consider a 
simpler problem in which the 
scale magnetic helicity is driven into the 
system quasi-isotropically
at the intermediate and small scales and the subsequent growth at 
at the large scale calculated.
This is  directly analogous to the  approach used to 
study nonlinear principles of velocity driven helical dynamos, and 
has led to significant progress in reconciling
results from analytic treatments and numerical
simulations.$^{\cite{pfl,b2001,bf02,maronblackman}}$
Keeping the set-up simple (a closed  system
driven quasi-isotropically with  kinetic or magnetic helicity
fluctuations), facilitates calculation of how much relaxation 
 occurs on dynamical vs. dissipative time scales, 
the strength of the backreaction on the relaxation
from velocity fluctuation growth, and the properties
of the saturated state.

Toward this end, I derive and solve a set of equations
that dynamically models the time evolution of a system
injected with bihelicity (opposite signs of magnetic helicities
on different scales) to determine the extent of the  magnetic helicity
generation on even larger scales.  The system is  magnetically
dominated in the sense that the energy injected into the system
is initially entirely magnetic, and the total 
magnetic energy (summed over all scales) remains 
dominant throughout the calculations
even when the velocity fluctuations grow. 
The time dependent evolution requires a three-scale nonlinear theory.
The  nonlinearity arises because
the magnetic helicity equations are coupled to 
dynamically to  equations for the velocity fluctuations.
The latter are generated
by the magnetically dominated system on its journey to the relaxed state.  


In Sec. II, I derive the equations and collect them in dimensionless form. 
In Sec. III, I solve the equations 
and discuss the solutions for the following cases: 
(1) driven bihelical magnetic relaxation,
where the system is continuously driven with 
oppositely signed magnetic helicities on different scales;
(2) kinetic helicity driven dynamo, where the system
is continuously driven with kinetic helicity only on the intermediate scale;
(3) unihelical magnetic relaxation, where the system is continuously driven
with 1 sign of magnetic helicity only on the intermediate scale;
(4) free bihelical magnetic relaxation, where the system is initially
injected with oppositely signed magnetic helicities on the intermediate
and small scales but with no subsequent driving. 
In Sec. IV, I summarize the results and the implications.

\section*{II. Equations for  bihelical magnetic relaxation}

In this section, the generalized set of
equations that can describe both kinetic helicity
and magnetic helicity driven dynamos is derived.

\subsection*{A. Magnetic helicity equations}

I  assume that there
are 3 scales on which magnetic helicity can 
reside,
defined such that their wave numbers
satisfy $k_3> k_2>k_1>0$.
I take $\bbB$, $\bbA$, and $\bbE$ 
to be the components of the magnetic field $\bfB$,  
vector potential $\bfA$,  and electric field $\bfE$ 
that vary on the  scale $k_1^{-1}$. 
The respective
quantities on scales $k_2^{-1}$ and $k_3^{-1}$ will be labeled
with subscripts $2$ and $3$ respectively.

In a closed or global 
volume, the total magnetic helicity, $H=\lb\bfA\cdot\bfB\rb$, 
satisfies$^{\cite{moffatt},\cite{woltjer},\cite{berger84}}$
\beq
\partial_t H=-2\lb{\bfE}\cdot\bfB\rb,
\label{helcons}
\ee
where $\bf E= -\partial_t {\bf A} -\nabla \phi$,
$\phi$ is the scalar potential, and 
$\lb\rb$ indicates a global volume average.
Such a conservation formula  also 
applies separately to each scale. That is, 
using $\bbE = -\partial_t \bbA -\nabla{\overline \phi}$, 
$\bfe_2 = -\partial_t \bfa_2 -\nabla\phi_2$
and $
\bfe_3= -\partial_t \bfa_3 -\nabla\phi_3$
and dotting with $\bbB$, $\bfb_2$ and $\bfb_3$ respectively
we  obtain
\beq
\partial_tH_1= -2\lb\bbE \cdot\bbB\rb
\label{H1}
\ee
\beq
\partial_tH_2=-2\lb{\bfe_2}\cdot\bfb_2\rb
\label{3}
\ee
and
\beq
\partial_tH_3=-2
\lb{\bfe_3}\cdot\bfb_3\rb,
\label{4}
\ee
%
where $H_1\equiv\lb\bbA\cdot\bbB\rb$, 
$H_2\equiv\lb\bfa_2\cdot\bfb_2\rb$, 
and  $H_3\equiv\lb\bfa_3\cdot\bfb_3\rb$.
The  expressions for 
$\lb\bbE\cdot\bbB\rb$,
$\lb{\bfe_2}\cdot\bfb_2\rb$, and $\lb{\bfe_3}\cdot\bfb_3\rb$
are  
\beq
\lb{\bbE}\cdot\bbB\rb =
-\lb{\emfb}\cdot\bbB\rb - \lb{\bfepo}\cdot\bbB\rb 
+\lambda\lb\bbJ\cdot\bbB\rb
\label{1b}
\ee
\beq
\lb{\bfe_2}\cdot\bfb_2\rb =\lb\emfb
\cdot\bbB\rb-\lb{\bfep}_2\cdot\bfb_2\rb + \lambda\lb \bfj_2\cdot \bfb_2\rb 
\label{2b}
\ee
and
\beq
\lb{\bfe_3}\cdot\bfb_3\rb =
\lb\bfep_2\cdot\bfb_2\rb+
\lb\bfepo\cdot\bbB\rb 
+ \lambda\lb \bfj_3\cdot \bfb_3\rb,
\label{3bb}
\ee 
where $\lambda$ is the magnetic diffusivity, 
$\bbJ\equiv\curl \bbB,\ \bfj_2\equiv\curl\bfb_2$ and $\bfj_3\equiv\curl\bfb_3$ 
are the normalized currents on the respective
scales defined in convenient units, $\emfb\equiv{\overline{\bfv_2\ts\bfb_2}}$,  $\bfepo\equiv{\overline{\bfv_3\ts\bfb_3}}$, 
$\bfep_2 \equiv(\bfv_3\ts\bfb_3)_2$, 
and $()_2$ indicates averages which vary on scales
 $\ge k_2^{-1}$ and $\bfv_2$ and $\bfv_3$ are the associated
velocities.

Combining equations (\ref{H1}-\ref{4}) with
(\ref{1b}-\ref{3bb}) gives
\beq
\partial_t H_1=
2\emf_{||}|H_1^{1/2}|k_1^{1/2}
+2\epo_{||}|H_1^{1/2}|k_1^{1/2}
-2\lambda k_1^2 H_1
\label{13b}
\ee
\beq
\partial_t H_2=
-2\emf_{||}|H_1^{1/2}|k_1^{1/2}
+2\ep_{2||}
H_2^{1/2}|k_1^{1/2}
-2\lambda k_2^2 H_2
\label{13bb}
\ee
and
\beq
\partial_t H_3=-2\epo_{||}|H_1^{1/2}|k_1^{1/2}
-2\ep_{2||}
H_2^{1/2}|k_1^{1/2}-2\lambda k_3^2 H_3,
\label{14b}
\ee
where 
$\emf_{||}\equiv\emfb\cdot{\bbB\over|\bbB|}$,
$\epo_{||}\equiv\bfepo\cdot{\bbB\over|\bbB|}$,
and $\ep_{2||}\equiv\bfep_2\cdot{\bfb_2\over|\bfb_2|}$.
Here the three-scale approach has been used to write
$\lb\bbJ\cdot\bbB\rb =k_1^2H_1$,
$\lb\bfj_2\cdot\bfb_2\rb =k_2^2H_2$,
and 
$\lb\bfj_3\cdot\bfb_3\rb =k_3^2H_3$, 
and 
maximal helicity has been assumed so that 
$\lb\bbB^2\rb = k_1^2H_1$, 
$\lb\bfb_2^2\rb = k_2^2H_2$,
$\lb\bfb_3^2\rb = k_3^3H_3$.


\subsection*{B. Time-dependent electromotive force equations}

A dynamical nonlinear helical dynamo theory
should include$^{\cite{kleeorin82},\cite{bf02}}$  
the time evolution of $\emfb$ and by generalization
to three scales, also the time evolution for $\bfepo$, 
and $\bfep_2$.
Of primary interest are the specific components 
$\emf_{||}$, $\epo_{||}$ and  $\ep_{2||}$.  We therefore have 
\beq
\partial_t \emf_{||}=\lb\partial_t\bfv_2\ts \bfb_2\rb\cdot{\bbB\over |\bbB|} 
+\lb\bfv_2\ts \partial_t\bfb_2\rb\cdot {\bbB\over |\bbB|} +\emfb\cdot\partial_t{\bbB\over |\bbB|}, 
\label{12a}
\ee
\beq
\partial_t \epo_{||}=\lb\partial_t\bfv_3\ts \bfb_3\rb\cdot{\bbB\over |\bbB|} 
+\lb\bfv_3\ts \partial_t\bfb_3\rb\cdot {\bbB\over |\bbB|} +\bfepo\cdot\partial_t{\bbB\over |\bbB|} 
\label{12a3}
\ee
and
\beq
\partial_t \ep_{2||}=(\partial_t\bfv_3\ts \bfb_3)_2\cdot {\bfb_2\over |\bfb_2|} 
+(\bfv_3\ts \partial_t \bfb_3)_2\cdot {\bfb_2\over |\bfb_2|} +
\bfep_2\cdot \partial_t{\bfb_2\over |\bfb_2|} 
\label{13a}
\ee
To proceed, we need equations for $\partial_t\bfb_2$, $\partial_t\bfb_3$, 
$\partial_t\bfv_2$, and 
$\partial_t\bfv_3$, which come from the 
 magnetic induction equation 
\beq
\partial_t\bfB=\curl (\bfv\ts \bfB) +\lambda \nabla^2\bfB, 
\label{ind}
\ee
and the incompressible Navier-Stokes equation
\beq
\partial_t\bfv= \bfv\ts \bfw -\nabla (P+\bfv^2/2) +\bfJ\ts\bfB +\nu \nabla^2\bfv,
\label{nav}
\ee
where $P$ is the thermal pressure, and $\bfw\equiv\curl \bfv$.
Because there are multiple fluctuation scales ($k_2^{-1}$ and $k_3^{-1}$),
we must perform multiple averaging.
To obtain the equations for quantities with subscript 2 and 3 note
that  
\beq
\partial_t\bfb_2+\partial_t\bfb_3= (\partial_t\bfB -\partial_t\bbB),  
\ee
so that 
\beq
\partial_t\bfb_2= (\partial_t\bfB -\partial_t\bbB)_2, 
\label{15}
\ee
and 
\beq
\partial_t\bfb_3= (\partial_t\bfB -\partial_t\bbB)-(\partial_t\bfB -\partial_t\bbB)_2.
\label{16}
\ee
The analogous relations apply for $\partial_t \bfv_2$ and $\partial_t\bfv_3$,
with the simplification that $\partial_t\bbV= 0$ for 
the present calculation. Thus
\beq
\partial_t\bfv_2+\partial_t\bfv_3= \partial_t\bfv ,  
\ee
so that 
\beq
\partial_t\bfv_2= (\partial_t\bfv)_2, 
\label{15v}
\ee
and 
\beq
\partial_t\bfv_3= \partial_t\bfv -(\partial_t\bfv)_2.
\label{16v}
\ee

After some algebra, applying this procedure to 
(\ref{ind}) and (\ref{nav})
gives 
\beq
\partial_t \bfb_2=\curl(\bfv_2 \ts \bbB) + \curl  (\bfv_2\ts \bfb_2)_2
+\curl(\bfv_3\ts \bfb_3)_2 +\lambda \nabla^2\bfb_2 + {\bf f}_{b2},
\label{17}
\ee

\beq
\partial_t \bfb_3=\curl(\bfv_2\ts\bfb_3)+\curl(\bfv_3\ts\bfb_2)+\curl(\bfv_3\ts\bbB)+\curl(\bfv_3\ts\bfb_3)_3 +\lambda\nabla^2\bfb_3 
+ {\bf f}_{b3},
\label{18}
\ee

\beq
(\partial_t \bfv_2)_i= \Pi_{ij}
(\bbB\cdot\nabla\bfb_2+\bfb_2\cdot\nabla\bbB)_j+
\nu (\nabla^2\bfv_2)_i + {\bf f}_{v2},
\label{19}
\ee
and 
\beq
(\partial_t \bfv_3)_i=\Pi_{ij}(\bbB\cdot\nabla\bfb_3+\bfb_3\cdot\nabla\bbB+
\bfb_2\cdot\nabla\bfb_3+\bfb_3\cdot\nabla\bfb_2)_j
+\nu (\nabla^2\bfv_3)_i+ {\bf f}_{v3},
\label{20}
\ee
where $\nu$ is the viscosity, 
${\bf f}_{b2}$ and ${\bf f}_{b2}$,
are forcing functions assumed not to correlate with $\bfv_1,\bfv_2$, 
and ${\bf f}_{v2}$ and ${\bf f}_{v3}$ are forcing functions assumed 
not to correlate with $\bfb_2,\bfb_3$. 
The projection operator $\Pi_{ij}\equiv\left(\delta_{ij}-{\nabla_i\nabla_j\over \nabla^2}
\right)$  comes from taking
the divergence of the velocity equation to solve for the 
total pressure of incompressible flow.

In deriving (\ref{15}-\ref{20}) and in subsequent manipulations, 
the following important simplifying
assumptions are made: (1) Mixed correlations between quantities of 
widely separated scales are assumed small. Thus I ignore terms
of the form e.g. $\lb\bfj_2\cdot\bfb_3\rb$.
(Note however that e.g. $\bfj_2\cdot\bfb_3\ne 0$; only
when put in brackets is the assumption applied.)
(2) The magnetic field or velocities on a given scale are assumed 
to be fully helical, so that specifically  
$\bfv_3\ts\bfw_3\simeq\bfv_2\ts\bfw_2\simeq
\bbV\ts(\curl\bbV)\simeq\bfj_3\ts\bfb_3\simeq\bfj_2\ts\bfb_2\simeq
\bbJ\ts\bbB\simeq 0$. This is not as restrictive as it might 
appear because the mixed quantities  e.g. $\bfj_2\ts\bfb_3$ are not assumed
to vanish and thus non-linear couplings and transfer between scales
are included.  
(3) Cross helicities are ignored (e.g. pseudo-scalar mixed correlations
between $\bfv$ and $\bfb$ such as $\lb\bfb\cdot\bfvp\rb$).
(4) It is assumed that terms involving the correlation of
products of separate quantities averaged over the intermediate scale can
be factored into two global correlations. That is,   
$\lb(\bfj_3\cdot\bfb_3)_2b_2^2\rb\simeq
\lb\bfj_3\cdot\bfb_3\rb \lb b_2^2\rb$
and $\lb(\bfv_3\cdot\bfw_3)_2b_2^2\rb\simeq
\lb\bfv_3\cdot\bfw_3\rb \lb b_2^2\rb$.

Armed with these assumptions, 
Eqs. (\ref{17}-\ref{20}) must now be used in Eqs. (\ref{12a}) and 
(\ref{13a}). Eqs. (\ref{17}) and (\ref{18}) are
required for the third terms in (\ref{12a}) and (\ref{13a}). 
These terms can be computed in configuration
space by direct substitution of the latter two equations into the former.
Eqs. (\ref{19}) and (\ref{20}) are required for 
the second terms in (\ref{12a}) and (\ref{13a}).
These terms are computed by Fourier transforms,$^{\cite{gd95},\cite{bf02}}$ 
and triple correlations are also retained by use of 
the ``minimal $\tau$'' closure $^{\cite{bf02}}$ 
(which obviates the 
first order smoothing approximation) by approximating the triple 
correlations as the product of the respective electromotive
force divided by a damping time scale. Applying this closure 
here, the triple correlation arising from the fifth term of
(\ref{18}), when input into the third term of (\ref{13a}), 
contributes  a term of the form $-\epo_{||} |v_3k_3|$, and
the third and fourth terms of (\ref{17}) when plugged into the
third term of (\ref{12a})  contribute terms of the form
$-\emf_{||} |v_2k_2|$, and $-\epo_{||} |v_3k_3|$ respectively.
It should also be noted that the last terms of (\ref{12a}) and
(\ref{13a}) respectively will also be included in these damping terms. 
The coefficients of the damping terms can in principle 
be taken to be different from unity$^{\cite{bf02}}$ 
but this is not considered further here.

Using the above approximations and 
plowing through the algebra that arises from plugging Eqs.  
(\ref{17}-\ref{20}) into  (\ref{12a}-\ref{13a}) gives 
\beq
\partial_t \emf_{||}=\left(k_2^2H_2-H_2^V\right){|H_1^{1/2}|k_1^{1/2}\over 3}
-{H_2^V \over 3}
{k_1^{3/2}\over k_2} {H_1\over |H_1^{1/2}|}
-\emf_{||}k_2^2(v_2/k_2+\lambda+\nu)-\ep_{2||}k_2v_2,
\label{E2}
\ee

\beq
\partial_t \epo_{||}=\left(k_3^2H_3-H_3^V\right){|H_1^{1/2}|k_1^{1/2}\over 3}
-{H_3^V\over 3}
 {k_1^{3/2}\over k_3} {H_1\over |H_1^{1/2}|}
-\epo_{||}k_3^2(v_3/k_3+\lambda+\nu),
\label{E3}
\ee
and
\beq
\partial_t \ep_{2||}=\left(k_3^2H_3-H_3^V\right){|H_2^{1/2}|k_2^{1/2}\over 3}
-{H_3^V\over 3}
 {k_2^{3/2}\over k_3} {H_2\over |H_2^{1/2}|}
-\ep_{2||}k_3^2(v_3/k_3+\lambda+\nu),
\label{e33}
\ee
where $H_2^V\equiv \lb\bfv_2\cdot\bfw_2\rb$ 
and $H_3^V\equiv \lb\bfv_3\cdot\bfw_3\rb$ 
are the kinetic helicities associated with the respective scales.

\subsection*{C. Kinetic helicity equations}

To close the set of equations, we need dynamical equations
for 
$H_2^V$ and $H_3^V$.
Using (\ref{15v}) and (\ref{16v}) in (\ref{nav}) to obtain
equations for $\partial_t \bfv_2$, $\partial_t \bfw_2$,
$\partial_t \bfv_3$ and $\partial_t \bfv_3$, ignoring
total divergences inside of averaging brackets and using
the assumptions discussed below (\ref{20}) 
we obtain
\beq
\partial_t H_2^V = 2\lb\bbJ\cdot(\bfb_2\ts\bfw_2\rb
+2 \lb\bbB\cdot (\bfw_2\ts\bfj_2)\rb
\simeq 2(k_2^2-k_1k_2)\lb\emfb\cdot \bfB\rb-2k_2^2\nu H_2^V
\label{26}
\ee
and
\beq
\begin{array}{r}
\partial_t H_3^V = 2\lb\bbJ\cdot(\bfb_3\ts\bfw_3\rb
+2 \lb\bbB\cdot (\bfw_3\ts\bfj_3)\rb 
+2 \lb\bfj_2\cdot (\bfb_3\ts\bfw_3)\rb 
+2 \lb\bfb_2\cdot (\bfw_3\ts\bfj_3)\rb \\
\simeq2(k_3^2-k_1k_3)\lb\bfepo\cdot \bbB\rb+
2(k_3^2-k_2k_3)\lb\bfep_2n\cdot \bfb_2\rb
-2k_3^2\nu H_3^V,
\end{array}
\label{27}
\ee
where the latter similarities in (\ref{26}) and (\ref{27})
follow from maximally helical assumptions  
$\bfj_2 \propto \bfb_2$,
$\bfw_2 \propto \bfv_2$,
$\bfj_3 \propto \bfb_3$, and
$\bfw_3 \propto \bfv_3$,
and the assumptions that 
${\bfw_2\cdot \bfv_2\over |\bfw_2||\bfv_2|}=
{\bfj_2\cdot \bfb_2\over |\bfj_2||\bfb_2|}$
and
${\bfw_3\cdot \bfv_3\over |\bfw_3||\bfv_3|}=
{\bfj_3\cdot \bfb_3\over |\bfj_3||\bfb_3|}$.
These latter relations will prove to be self-consistent
for the solutions discussed later.

\subsection*{D. Aggregate set of equations in dimensionless form to be solved}

Before collecting the complete set of equations to be solved, 
it is useful to add injection and loss 
terms to the three magnetic helicity 
equations derived above.  Since we will study 
the evolution of magnetic helicity 
for a system subject to injection of 
bihelicity---that is, helicity injected
of opposite sign on the scales $k_2^{-1}$ and $k_3^{-1}$---we 
add terms to the $H_2$ and $H_3$ equations that account for 
injection (or loss) of magnetic helicity on these respective scales.
In addition, if the system is a corona of finite volume,
large scale fields therein might be able to escape as Poynting flux. 
While the injection terms just described can be used to 
combine  injection and loss for $H_2$ and $H_3$,
for $H_1$ I add a distinct Poynting loss term 
$\propto {\bbV}_A H_1k_1 \sim H_1^{3/2}k_1^{1/2}$, 
where $\bbV_A=\bbB$ is the Alfv\'en
speed associated with $\bbB$ (recall that magnetic fields are 
written in Alfv\'en units herein).  
(When losses are included, the magnetic helicities should 
be re-interpreted as gauge invariant relative helicities$^{\cite{berger84}}$
with respect to a background potential field, but 
the basic formalism herein does not change significantly.)

For completeness, I also include 
injection (or loss) terms 
in the $H_2^V$ and $H_3^V$ equations.
Use of the injection term for $H_2^V$ will help
illustrate the relation of  the present study
to the 2-scale solution of the kinetic helicity driven dynamo.$^{\cite{bf02}}$

To write the needed   equations 
in dimensionless form, define 
$h_1\equiv H_1 /|{\tilde H}_{2,0}|$, 
$h_2\equiv H_2 /|{\tilde H}_{2,0}|$, 
$h_3\equiv H_3 /|{\tilde H}_{2,0}|$,
$h_2^V\equiv H_2^V /|k_2^2 {\tilde H}_{2,0}|$, 
$h_3^V\equiv H_3^V /|k_2^2 {\tilde H}_{2,0}|$, 
$R_M\equiv {\tilde H}_{2,0}^{1/2}/ \lambda k_2^{1/2}$, 
$R_V\equiv {\tilde H}_{2,0}^{1/2}/ \nu k_2^{1/2}$, 
$\tau \equiv t k_2^3/2 {\tilde H}_{2,0} $, 
${\overline Q}=\emf_{||}/k_2{\tilde H}_{2,0}$, 
${\overline q}=\epo_{||}/k_2{\tilde H}_{2,0} $,  
and $q_2=\ep_{2||}/k_2{\tilde H}_{2,0}$, 
where ${\tilde H}_{2,0}=H_2(0)$ if
magnetic helicity is initially injected 
and ${\tilde H}_{2,0}=H_2^V(0)/k_2^2$ if
kinetic helicity is initially injected.
Using these dimensionless parameters and adding 
the afore mentioned injection and loss terms,  
Eqs. (\ref{13b}-\ref{14b}) become
\beq
\begin{array}{r}
\partial_\tau h_1 = {2}{\overline Q}|h_1^{1/2}|
\left({k_1\over k_2}\right)^{1/2}
+2{\overline q}|h_1^{1/2}|\left({k_1\over k_2}\right)^{1/2}
-{2h_1\over R_m}\left({k_1\over k_2}\right)^2-
c_1|h_1^{3/2}|(k_1/k_2)^{3/2},
\end{array}
\label{h1}
\ee

\beq
\begin{array}{r}
\partial_\tau h_2 = s_2
-2{\overline Q}|h_1^{1/2}|
\left({k_1\over k_2}\right)^{1/2}
+2q_{2}|h_2^{1/2}|-{2h_2\over R_m},
\end{array}
\label{h2}
\ee
and
\beq
\begin{array}{r}
\partial_\tau h_3 = s_3
-2{\overline q}|h_1^{1/2}|\left({k_1\over k_2}\right)^{1/2}
-2q_{2}|h_2^{1/2}|
-{2h_3\over R_m}\left({k_3\over k_2}\right)^{2},
\end{array}
\label{h3}
\ee
where $c_1$ is a constant parameterizing the Poynting loss
term and $s_1$ and $s_2$ are the helicity injection terms
described  in the previous paragraph.

Completing the set of equations using the 
dimensionless variables defined above, we have for 
Eqs. (\ref{E2}-\ref{e33}) and (\ref{26}-\ref{27})
\beq
\partial_\tau {\overline Q}=
{1\over 3}(h_2-h_2^V)|h_1^{1/2}|\left({k_1\over k_2}\right)^{1/2}
-{1\over 3}|h_2^V|\left({h_1\over |h_1^{1/2}|}\right)
\left({k_1 \over k_2}\right)^{3/2}-{\overline Q}\left(|\sqrt{h_2^{V}}|+{1\over R_M}+{1\over R_V}\right) -q_{2}|\sqrt{h_2^{V}}|,
\label{Q2}
\ee

\beq
\partial_\tau {\overline q}=
{1\over 3}(h_3-h_3^V)|h_1^{1\over 2}|\left({k_1\over k_2}\right)^{1\over 2}\left({k_3\over k_2}\right)^{2}
-{1\over 3}|h_3^V|\left({h_1\over |h_1^{1\over 2}|}\right)
\left({k_1\over k_3}\right)\left({k_1\over k_2}\right)^{1\over 2}
-{\overline q} {k_3^2\over k_2^2}
\left({k_2\over k_3}|{\sqrt{h_3^{V}}}|+{1\over R_M}
+{1\over R_V}\right), 
\label{Q3}
\ee

\beq
\partial_\tau q_{2}=
{1\over 3}(h_3-h_3^V)|h_2^{1/2}|\left({k_3\over k_2}\right)^{2}
-{1\over 3}|h_3^V|\left({h_2\over |h_2^{1/2}|}\right)
\left({k_2\over k_3}\right)
-q_{2} {k_3^2\over k_2^2}
\left({k_2\over k_3}|{\sqrt{h_3^{V}}}|+{1\over R_M}
+{1\over R_V}\right),
\label{q3}
\ee

\beq
\begin{array}{r}
\partial_\tau h_2^V= s_2^V+2{\overline Q}|h_1^{1/2}|\left({k_1\over k_2}\right)^{1/2} 
\left(1-{k_1\over k_2}\right)-{2h_2^V\over R_V},
\end{array}
\label{h2v}
\ee
and
\beq
\begin{array}{r}
\partial_\tau h^V_3= s_3^V + 2\left({k_3^2\over k_2^2}-{k_1k_3\over k_2^2}\right)
{\overline q}|h_1^{1/2}|\left({k_1\over k_2}\right)^{1/2}
+
2\left({k_3^2\over k_2^2}-{k_3\over k_2}\right)q_{2}|h_2^{1/2}|
-{2h_3^V\over R_V}\left({k_3\over k_2}\right)^{2},
\end{array}
\label{h3v}
\ee
where $s_2^V$ and $s_3^V$ are the analogous source
terms to $s_2$ and $s_3$. 
Eqs. (\ref{h1}-\ref{h3v}) 
 can be solved for $h_1$, $h_2$, 
$h_3$, ${\overline Q}$, ${\overline q}$, $q_{2}$, 
$h_3^V$, and $h_2^V$ as a function of time for
a variety of different initial conditions.
A few cases are studied below.

\section*{III. Discussion of solutions}

Eqs. (\ref{h1}-\ref{h3v})  provide a unifying framework
for both velocity driven and magnetically driven nonlinear dynamos.
They characterize the magnetic and kinetic helicity dynamics for 
a range of  different physical scenarios, depending
on the initial conditions. 
In all cases, below $k_1=1$, $k_2=5$ and $k_3=20$.

\subsection*{A. Driven dynamical bihelical magnetic relaxation}

Consider the case 
for which $\partial_t h_3=\partial_t h_2=s_2^V=s_3^V=0$
and $h_2=-h_3=1$ for all times.
This implies continuous injection and removal of oppositely signed
magnetic helicities on the intermediate and small scales
so as to maintain $h_2=-h_3$.
Fig. 1 shows the solutions of  
Eqs. (\ref{h1}-\ref{h3v}) for initial $(t=0)$ values
$h_1(0)=0.01$,  $h_2^V(0)=h_3^V(0)={\overline Q}(0)=
{\overline q}(0)=q_2(0)=0$.
Fig 1a shows  $h_1$ through the saturated regime and  
Fig 1b focuses on its early time solution.
Fig 1c shows $h_2$ and $h_2^V$, Fig 1d shows $h_3$ and $h_3^V$,
Fig.1e shows ${\overline Q},{\overline q},$ and $q_2$,
and Fig 1f is a diagnostic solution. 
These are discussed below.

\subsubsection*{1. Late time regime}

To analytically estimate the late time steady state solution for $h_1$
of Fig 1a, note that from equation (\ref{h2v}) in the steady state, 
${\overline Q}$ is inversely proportional to the large quantities 
$R_V$ and $h_1^{1/2}$.  Assuming that 
$|{\overline Q}|<<|{\overline q}|$  and $|{\overline Q}|<<| q_{2}|$ 
(these will turn out to be valid assumptions), 
we can then use Eqs. (\ref{h1}), (\ref{q3}), and (\ref{h3v}) to 
find the saturation value of  $h_1$. 

In the steady-state, Eq. (\ref{h3v}) gives
\beq
q_{2}\simeq -{\overline q}h_1^{1/2}\left({k_1\over k_2}\right)^{1/2}\left({1-k_1/k_2 \over 1-k_2/k_3}\right),
\label{3.31}
\ee
assuming that the last term in (\ref{h3v}) is small.
The fact that $q_2$ asymptotes to  negative 
values implies that  $h_3\sim h_3^V$ in (\ref{q3}) 
due to the growth of $h_3^V$.
The steady limit of Eq. (\ref{q3}) 
can then be rearranged to give the approximate relation
\beq
q_{2}\simeq -{1\over 3}{k_2^2 \over k_3^2},
\label{3.32}
\ee
where we have neglected the $R_M$ and $R_V$ terms in that equation as they
are small.
Combining (\ref{3.31}) and (\ref{3.32}) gives
\beq
{\overline q}
\simeq {1\over 3}{k_2^2 \over k_3^2}\left({h_2\over h_1}\right)^{1/2}
\left({k_2\over k_1}\right)^{1/2}\left({1-k_2/k_3 \over 1-k_1/k_3}\right).
\label{3.33}
\ee
This equation can then be used in the steady-state limit of Eq. (\ref{h1}).
Using the earlier assumption that 
$|{\overline Q}|<<|{\overline q}|$, we then have
\beq
h_1 \simeq {R_M\over 3}{k_2^4\over k_1^2k_3^2}\left({1-k_2/k_3\over1-k_1/k_3}\right),
\label{exc}
\ee
which provides  an excellent approximation to the 
actual solution at late times. It also shows
the insensitivity to $R_V$.

The saturated $h_1\propto R_M$ in (\ref{exc}),  
is also exhibited in Fig. 1a.  
This can be understood as follows:
$h_1$ grows first with the sign of $h_2$ by
direct analogy to unihelical magnetic relaxation,$^{\cite{bf04}}$ 
(see also sec. 3.2.2):  The magnetic helicity seeks the largest
scale available of a given sign.
However in the bihelical case,  the additional 
injection of negative $h_3$ drives the growth of negative $h_3^V$.
In fact, $|h_3^V|$ grows to very slightly exceed $|h_3|$ as seen
in Fig. 1d. Then, $\overline q$ becomes slightly 
positive.  This drives  additional $h_1$ growth by an induced kinetic helicity driven dynamo from $h_3^V$, by maintaining the 
$\overline q$ term on the right of (\ref{h1}) at a positive value
(long dashed curve in fig 1e).
The larger the value of $R_M$,  the less competitive
the last term on the right of  (\ref{h1}) is with the $\overline q$ growth term. This explains the $R_M$ dependence of the
saturation value of $h_1$.

The above interpretation can be tested by artificially 
reducing the coupling of $h_3^V$ to the other equations.
This was done to produce the solutions
in Fig. 1f by multiplying the entire
right side of (\ref{h3v}) by $10^{-6}$. 
Notice the significant reduction in the value
of $h_1$ at even relatively early  times compared to Fig 1a.  In this case,
the presence of $h_3$ acts to damp the growth of $h_1$ because
$\overline q$ remains negative at all times.
The two curves in Fig 1e also confirm that 
it is indeed the ${q_2}$ term in (\ref{Q2}) that 
leads to the rapid growth of $h_1$ in Fig 1a,b since
lowering the coefficient on that term lowers the saturated
$h_1$.

\subsubsection*{2. Early time, kinematic regime}

In Fig. 1b, the early time solution for $h_1$ is shown.
The solution is independent of $R_M$ up to $h_1\sim 1$.
At early times, growth of $h_1$ is dominated by
the first term on the right side of (\ref{h1}).
The dominant term for the growth of $h_2^V$ from (\ref{h2v})
is very similar and thus the two initially grow at nearly  the same
rate.  However, $h_2^V$ 
eventually builds up to a value that 
stops the growth of $\overline Q$ in (\ref{Q2}), 
after which the latter decays and the growth of 
$h_1$ and $h_2^V$ slows.  The decay of $\overline Q$ occurs
when the first two terms on the right of (\ref{Q2}) approximately
balance. 
Since $h_2=1$, the critical $h_2^V$ is $\sim h_2/(1+k_1/k_2)
\sim 0.83$. This provides an excellent match to the value of $h_2^V$ at the
end of the kinematic regime shown in Fig 1e. 
Since the growth rate 
of $h_2^V$ is slightly less than that of $h_1$ 
due the presence of the $(1-k_1/k_2)$  factor 
in the first term on the right
of (\ref{h2v}) compared to the first term on the  right of (\ref{h1}), 
$h_1$ grows a bit larger than $h_2\sim 1$ 
during the kinematic regime.  After this time the growth depends on $R_M$.
Just after the kinematic regime, the $h_1$ curve for lower
$R_V$ is the higher of the two curves in Fig. 1b. This is because
a lower $R_V$ reduces the kinetic helicity backreaction
in $\overline Q$ since the velocity decays faster.
However, the two curves eventually cross to produce the
late time solution of Fig. 1a for the reasons
discussed in the previous subsection.

The value of $h_1$ reached during the kinematic regime during
bihelical relaxation is approximately 
the same as that reached in the kinematic regimes of
the simpler two-scale unihelical magnetic relaxation model$^{\cite{bf04}}$
and the kinetic helicity driven helical dynamo$^{\cite{bf02}}$ for the 
similar reason that 
in these cases, growth of the largest scale helicity 
is driven by the difference between magnetic and kinetic helicities
from the small scale. Driving a two scale  system with kinetic (magnetic)
or magnetic helicity grows the large scale $h_1$ rapidly at first,
but then the magnetic (kinetic) helicity on the small scale
builds up and quenches the growth to dissipation limited rates.
As discussed in the previous subsection, 
the evolution of $h_1$ differs for bihelical relaxation
long after the kinematic regime. Sec. III.B shows how  all of these cases 
are consistent and just depend on the choice of initial conditions.

Finally, a few more comments on Fig. 1c and d:
Although curves for both $R_M=1000$ and $R_M=6000$ 
are shown in both panels, they overlap at early times 
when there is  no  $R_M$ dependence.  Also notice the oscillations
in $h_3^V$, which damp very quickly, after which $h_3^V$ remains steady.
Remember that the unit of time is defined on the
$k_2^{-1}$ scale. The short evolution time of $h_3^V$
results  from the factors of $(k_3/k_2)^2=16$
in Eq. (\ref{h3v}).

\subsubsection*{3. Comparing $k_2=k_3$ vs $k_3 > k_2$}

If $k_2=k_3$, then for the case of Fig. 1 where
$h_2=-h_3=1$,  the net injection of helicity at the
scale $k_2=k_3$ would be zero and nothing should happen.
This has been tested and indeed there is no growth for
this case. This highlights that although  growth
of $h_1$ can occur when the net injection of $h_2 + h_3=0$,
the two signs of helicity must be injected on sufficiently 
different scales ($k_3^2/k_2^2 >>1$) for non-trivial solutions. 

\subsubsection*{4. Inclusion of boundary terms}

In Fig.1, $c_1=0$.
Fig. 2 shows the effect of varying $c_1 >0$, to approximate
different finite loss rates of $h_1$.
The  result is that the value of $h_1$ saturates
at much lower values than in Fig. 1a, as  expected.

\subsubsection*{5. Case of fixed $\partial_t h_2=-\partial_t h_3 =$ constant}

In the previous subsections,  $s_2=s_3=0$ 
and $h_2=1$ and $h_3=-1$ were held constant.
The case of constant injection rate, namely $s_1=1$ and $s_2=-1$, 
is a natural alternative 
to consider.  Solutions for this case also reach a steady state
at late times. Higher values of $h_1$ are obtained in this case 
than for the fixed $h_2=-h_3$ case. I do 
not discuss this case further here as the implications are similar.

\subsection*{B. Kinetic helicity driven dynamo and driven unihelical relaxation}

%
In previous two-scale and three-scale approaches for the
kinetic helicity driven dynamo,$^{\cite{bf02}-\cite{b03}}$ dynamical
equations for the kinetic helicity 
were unnecessary because the kinetic helicity was kept
steady. In contrast, the unihelical relaxation study$^{\cite{bf04}}$
did include kinetic helicity dynamics but it 
was a two-scale approach with only 1 sign of injected magnetic helicity. That the equations (\ref{h1}-\ref{h3v}) are more general
means that they should reduce to these other cases in the appropriate
limits.

\subsubsection*{1. Kinetic helicity driven dynamo}

If the system is driven steadily with $h_2^V$ 
starting with only a seed $h_1$ and 
$h_2=h_3\simeq 0$, 
then the solution reproduces the non-linear multi-scale helical 
dynamo$^{\cite{b03}}$, which in turn is consistent with the results of the two-scale
version.$^{\cite{bf02}}$  
Fig. 3 shows the resulting solution for $h_1$, which represents a simple
$\alpha^2$ dynamo.$^{\cite{b2001,bf02}}$  
The rate of growth up to $h_1=1$ is 
fast and independent of $R_M$. The ultimate steady
state ratio $h_1/h_2=(k_2/k_1)^2 =25$ 
to which the solution converges at late times ($t>R_M k_2/k_1$),
is also independent of $R_M$. This can be seen from solving
equations (\ref{h1}) and (\ref{h2}) with $s_2=c_1={\overline q}=q_2=0$
in the steady state).  

The rate of approach to the steady state is slower for smaller $R_M$.
This is because the growth of $h_1$ for the 
$h_2^V$ driven dynamo occurs by  segregating 
opposite signs of magnetic helicity 
to small and large scales. 
Growing $h_1$ larger than $h_2$ means that  $h_2$ has
to dissipate faster than $h_1$. The larger the $R_M$, the slower
the dissipation of $h_2$, leading to a smaller $\overline Q$, 
and a longer time  for $h_1$ to saturate.

This late-time dependence on $R_M$ differs from that 
of $h_1$ growth in  bihelical magnetic relaxation 
as discussed in Sec. III.A.1. 
There, a larger $R_M$ gave a larger saturation value
and a faster growth rate at late times.  
The difference arises because growing $h_1$ at
late times in that case is not determined  by losing small scale 
magnetic helicity but  by maximizing the effect of 
the secondary induced $h_3^V$ driven dynamo.
There is no contradiction in the role of $R_M$ 
between these two cases because 
of the difference in the initial conditions.

The saturated value of  $h_1$ depends  
slightly on the value of turbulent diffusion 
(the third term of (\ref{Q2})) 
and its quenching. 
Different scaling factors for turbulent diffusion were 
used previously$^{\cite{bf02,bb02}}$
and these differences explain 
the small difference in the saturation values therein when compared
to Fig 3. There is no dynamical quenching of turbulent diffusion
in either the present or previous approaches,
so here I simply eliminated the
extra multiplicative phenomenological factors.

In short, the equations of Sec. 2 fully account
for results in the previous studies of the nonlinear
$\alpha^2$ dynamo.


\subsubsection*{2. Unihelical magnetic relaxation}

If the system is driven with  $h_2$ but without $h_3$,
then the growth of $h_1$ responds just as in 
unihelical dynamical relaxation.$^{\cite{bf04}}$
The injected helicity  $h_2$ drives the growth
of $h_1$ by dynamical relaxation. (This is called unihelical
relaxation because the helicity is injected with 1 sign at $h_2$
and the same sign then grows at $h_1$.)
Fig. 4 shows the growth of $h_1$.
The kinematic regime proceeds until $h_1\sim 1$, after
which the growth of $h_1$ depends on dissipation.
However, unlike the bihelical magnetic relaxation of section 3.1,
here the late time 
saturation value of $h_1$ depends on the ratio $R_V/R_M$.  
Here $h_2^V$
plays the role of back reactor, and a larger $R_V$ means that
$h_2^V$ is more effectively drained,
allowing further $h_1$ growth until saturation.
Again note that in the contrasting  bihelical case of Fig. 1, 
the saturation value of $h_1$ is insensitive to $R_V$ (for $R_V>>1$).
There is no  contradiction because in that case, both positive
and negative signs of magnetic helicity are injected
and the $h_3=-1$ injection produces an additional $h_3^V$ 
driven dynamo through $\overline q$ in (\ref{h1}) that dominates
any effect of the suppression of $\overline Q$ at late times.

The important point is that the previously studied case of two-scale 
unihelical relaxation emerges from the more general equations of Sec. 2 
by the appropriate choice of initial conditions.

\subsection*{C. Free dynamical bihelical magnetic relaxation}

In the previous sections, either kinetic or magnetic helicity was 
assumed to be continuously driven into the system.
Here I consider the case where with initial $h_2(0)=-h_3(0)=1$
and $h_1(0)=0.01$, but with no subsequent driving or boundary terms, 
i.e. $c_1=s_2=s_3=s_2^V=s_3^V=0$. The system is then 
allowed to freely relax.  Results of this case are shown in Fig. 5a-f.

From Fig. 5a and 5b, it is evident that significant growth of $h_1$ occurs. 
The kinematic
phase (independent of $R_M$ and $R_V$) ends when $h_2\sim h_2^V\sim h_1\simeq 1/2$  by analogy to free unihelical relaxation,$^{\cite{bf04}}$
due to transfer of $h_2$ to $h_1$.
Fig. 5b shows that  $h_1\sim 1/2$ rather than $h_1\sim 1$ 
is reached in the kinematic phase 
because $h_2$ also drives the growth of $h_2^V$.  When $h_2^V\sim 1/2$, transfer of $h_2$ to $h_1$ is shut down as 
$\overline Q$ depletes
from the decrease in the first term on the right of (\ref{Q2}).
After the kinematic phase, the $h_3^V$ induced secondary dynamo
described in Sec. III.A takes over to amplify $h_1$ a bit beyond the kinematic
value. Here this has only a small effect because 
without forcing the system eventually decays.

For larger $R_M$, the slower the eventual
decay of $h_1$. This is as expected from (\ref{h1}), but with the subtle
implication that the ${\overline Q}$ and ${\overline q}$ 
terms combine  to give a net contribution that does not 
affect the sensitivity to $R_M$ from the dissipative term.
For fixed $R_M$ but smaller $R_V$, the decay rate of $h_1$ is the same,
but the maximum $h_1$ reached is smaller.  This 
is evident from comparing the  middle and top curves 
of Fig. 5a and is 
correlated with the faster depletion of $h_3^V$ and $h_3$ 
in Fig. 5c and 5e for the lower $R_V$ case.  
Remember that $h_3^V$ drives a secondary
dynamo that grows $h_1$ beyond the kinematic regime. As $R_V$ is
decreased, $h_3^V$ decays faster making 
this secondary dynamo is less effective.
Because $h_3$, and thus $h_3^V$ are not continuously supplied,
the secondary dynamo decays and so does $h_1$.
For arbitrarily larger $R_V$ than shown in Fig. 5a, 
the maximum reached by $h_1$ before it starts to decay is 
$h_1 \sim (1-k_2/k_3)=0.75 < h_2(0)$.

The curves for $h_3$ and $h_3^V$ indistinguishably overlap
in Fig. 5c but a blow up in Fig 5d shows that the $h_2^V$ curves
lie slightly to the right of the $h_3$ curves. This slight
offset is enough to drive the dissipation limited secondary dynamo
just described.  Fig. 5c also shows that $h_3^V$ and $h_3$ deplete faster for
smaller $R_M$. This is expected from Eqs. (\ref{h3})
and (\ref{h3v}) but with the implication  
that the $\overline q$ and $\overline q_2$ terms
combine to give a net contribution
that does not contradict the $R_M$ and $R_V$ 
dependence from the dissipative terms. 

In contrast to the slower decay of $h_1$, $h_3$ and $h_3^V$ with increased
$R_M$, Fig. 5b. shows that $h_2$ and $h_2^V$ actually decay faster
with increasing $R_M$. 
At first this may seem counterintuitive but
it results because  $q_2$ is negative and decays
more slowly with increasing $R_M$. Thus the contribution from the $q_2$ term
in (\ref{h2}) acts as a decay term. That this term is larger for
larger $R_M$ is evident from 
Fig. 5f where the bottom two curves show $q_2$ for $R_M=2000$ and
$R_M=10000$ respectively.  The bottom two curves are close
to the asymptotic values for large $R_M$. Little change 
occurs for higher value $R_M>10000$ (not shown).

Despite the subtleties, the key point to take away from the free
bihelical relaxation solution is that $h_1$ grows fast
and decays more slowly than $h_2$ or $h_3$.  Like the driven 
case of Sec. III.A, 
when the system is injected with magnetic helicity of equal magnitude but
opposite signs on intermediate and small scales 
$k_2^{-1}$ and $k_3^{-1}$,  
significant magnetic helicity grows on the largest $k_1^{-1}$
scale kinematically.
For injected values of $h_2=-h_3=1$,
$h_1$ grows to $h_1\sim h_2(0)/2\sim 1/2$ kinematically and 
independent of $h_1(0)$. The growth time is 
approximately the Alfv\'en crossing time of the $k_1^{-1}$ scale
using the injected field strength from $h_2$.
Unlike the driven case where the maximum $h_1$ reached
at late times is proportional to $R_M$, in the free case
the maximum post-kinematic $h_1$ reached for  $R_M\rightarrow \infty$ is 
$h_1\lsim 1$. Subsequently,  $h_1$ resistively decays.

The free  bihelical relaxation treated in this section is 
different from that of Ref. (\cite{yousef03}).  
There, the system was initially
injected with helicity of one sign on  the largest scale and the opposite
sign on a smaller 
scale. In contrast, here the helicity is injected
on two scales both smaller than the largest scale and helicities on 
all three scales are allowed to evolve.  Once $h_1$
grows to its maximum and $h_2$ depletes,  the subsequent resistive
evolution of $h_1$ and $h_3$ acts qualitatively like the  
relaxation of the random bihelical case of Ref. ({\cite{yousef03}) in that  
$h_1$ resistively decays more slowly than $h_3$.

\section*{IV. Summary and Implications}

A generalized set of equations that encompasses non-linear
velocity driven dynamos as well as magnetically driven dynamos (equivalently, 
dynamical magnetic relaxation) is
derived and solved herein. Previous work 
focused on these processes separately, but here 
the different processes are   unified
into a single set of equations that includes the dynamical
evolution of  kinetic helicity, magnetic helicity, 
and the turbulent electromotive forces. 
Particularly important is that the present approach 
includes the dynamics of 3 scales: two fluctuation scales
in addition to the  global scale, rather than just a single fluctuation
scale and global scale. This is fruitful because 
it allows studying bihelical relaxation, 
the evolution of the helicities on all three scales 
when helicities of opposite 
signs are injected on the intermediate and  small scale 
respectively. 

Example solutions of the bihelical magnetic relaxation equations 
reveal that even when a net zero magnetic helicity is supplied
to the system by injecting equal and opposite values 
on sufficiently separated intermediate and small scales, 
the large scale magnetic helicity  grows kinematically 
(independent of the magnetic Reynolds number)
to approximately
equal the sign and magnitude of the helicity injected on the intermediate 
scale.  
The oppositely signed magnetic helicity
injected on the  small scale subsequently drives a kinetic helicity
driven dynamo through the secondary amplification of kinetic helicity.
For the driven case, in which the magnetic helicity injection is continuous,
$h_1$ grows beyond its
kinematic value to a late time saturation value proportional 
to $R_M$ for a closed or infinite system. 
For the free relaxation case, in which the driving is turned off, 
the secondary dynamo growth 
 amplifies $h_1$ only slightly beyond its kinematic growth value of $h_2(0)/2$, 
to a maximum value that is independent of $R_M$ and $\le h_2(0)$. 
A real astrophysical corona likely never reaches the 
$R_M$ dependent regime as $R_M$ is extremely large in practice. 
The  magnitude of the large scale field growth 
taken from the end of the kinematic regime is then the most relevant.
For bihelical injection of equal and opposite signs
on the $k_2$ and $k_3$ scales, both injected with maximally helical
fields, the ratio of magnetic energy on
the large scale to that of the intermediate scale at the end of the 
kinematic regime is approximately $k_1/k_2$.

When the appropriate initial conditions are used,
the solutions of the generalized equations 
also reproduce the results of 
the previously studied simpler cases of 
unihelical magnetic relaxation$^{\cite{bf04}}$
and the kinetic helicity driven dynamo.$^{\cite{bf02,bb02}}$

In the context of magnetically dominated astrophysical coronae of 
stars or disks, bihelical magnetic helicity 
injection is natural because such coronae are fed with  fields
by velocity driven dynamos operating in their rotators 
below.  Nonlinear helical velocity driven dynamos 
produce helical fields of opposite signs on different 
scales$^{\cite{bf2000,bb03}}$ 
which can then emerge as bihelical fields into the corona.
(Shear, not considered here,  may effect the relative fraction injected
on each scale$^{\cite{vc01,bs04}}$.) 
The ``large'' scale and ``small'' scale of the rotator 
correspond to the ``intermediate'' and ``small'' scales of the corona
respectively. The corona is therefore  a natural environment for  
bihelical relaxation. The results herein suggest that 
helical large scale coronal fields can rapidly grow 
even when the coronae are injected with 
a net zero magnetic helicity. This is favorable for the in situ production
of fields mediating  coronal holes  or MHD outflows from stars and disks. 

Much more work is needed to incorporate
the basic concepts and idealized solutions 
developed herein into realistic models of coronae.
Shear, non-helical magnetic 
energy dynamics, and realistic boundary conditions should
all be included.


\ni 
Thanks to A. Brandenburg, G. Field, and
E. Vishniac for related discussions and comments.
DOE grant DE-FG02-00ER54600 
and NSF grant AST-0406799 are 
acknowledged. Thanks also to KITP of UCSB, where this research was supported in part 
by NSF grant PHY99-07949.


\eject
\enumerate

\bibitem{moffatt} H.K. Moffatt, {\sl Magnetic
Field Generation in Electrically Conducting Fluids}, 
(Cambridge University Press, Cambridge, 1978)

\bibitem{parker}  
E.N. Parker, {\it Cosmical Magnetic Fields}, (Oxford: Clarendon
Press, 1979)

\bibitem{krause}   F. Krause \& K.-H. R\"adler, 
{\it Mean-field Magnetohydrodynamics and Dynamo Theory}, 
(Pergamon Press, New York, 1980)

\bibitem{zeldovich83} 
Ya. B. Zeldovich , A.A. Ruzmaikin, \& D.D. Sokoloff, {\sl Magnetic Fields in Astrophysics}, 
(Gordon and Breach, New York, 1983)

\bibitem{strauss85} 
H.R. Strauss, Phys. Fluids, {\bf 28}, 2786 (1985)

\bibitem{strauss86} 
H.R. Strauss, Phys. Fluids, {\bf 29}, 3008 (1986)

\bibitem{ortolani93} 
S. Ortolani  \& D.D. Schnack, 
{\it Magnetohydrodynamics of Plasma Relaxation}
(World Scientific: Singapore, 1993)

\bibitem{bhattacharjee86} 
A. Bhattacharjee \& E. Hameiri,  Phys. Rev. Lett. 
{\bf 57}, 206 (1986)

\bibitem{holmesetal88} 
J.A. Holmes, B.A. Carreras, P.H. Diamond, \& V.E. Lynch,  
Phys. Fluids, {\bf 31}, 1166 (1988)

\bibitem{gd95}  
A.V. Gruzinov \& P.H. Diamond, Phys. Plasmas, {\bf 2}, 1941 (1995)

\bibitem{by}  A. Bhattacharjee \& Y. Yuan, Astrophys. J.,  
{\bf 449}, 739 (1995)

\bibitem{bellan00}  P.M. Bellan, {\sl Spheromaks}, 
(Imperial College Press, London, 2000)

\bibitem{bf04}E.G. Blackman and G. B. Field, 
Phys. Plasmas {\bf 11}, 3264 (2004)

\bibitem{jiprager02} 
 H. Ji \& S.C. Prager,  {Magnetohydrodynamics}
{\bf 38}, 191  (2002)

\bibitem{pfl}
A. Pouquet, U. Frisch,  J. L\'eorat,   J. Fluid Mech., {\bf 77}  321 (1976)

\bibitem{kleeorin82}
N.I. Kleeorin  \& A.A. Ruzmaikin, 
{Magnetohydrodynamics}, {\bf 2}, {17}, (1982)

\bibitem{bf2000} 
E.G. Blackman,  \& G.B. Field, Mon. Not. R. Astron. Soc. {\bf 318}, 724 (2000)

\bibitem{b2001}
A. Brandenburg, Astrophys. J., {\bf 550}, 824 (2001)

\bibitem{fb02} 
G.B. Field  \& E.G. Blackman, Astrophys. J., {\bf 572}, 685 (2002)

\bibitem{bf02} E.G. Blackman \& G.B. Field, 
Phys. Rev. Lett., {\bf 89}, 265007 (2002)

\bibitem{bb02}E.G. Blackman \& A. Brandenburg,
Astrophys. J. {\bf 579}, 359 (2002)

\bibitem{b03}
E.G. Blackman, Mon. Not. R. Astron. Soc., {\bf 344}, 707  (2003)

\bibitem{bb03} E.G. Blackman 
\& A. Brandenburg, Astrophys. J. Lett., {\bf 584} L99  (2003)

\bibitem{matthaeus04}
A. Brandenburg \& W.H. Matthaeus, Phys. Rev. E., {\bf 69}, 056407 (2004)

\bibitem{maronblackman}
J. Maron \& E.G. Blackman, Astrophys. J. Lett. {\bf 566}, L41 (2002)

\bibitem{bs2004}
A. Brandenburg \& K. Subramanian, ``Astrophysical magnetic fields and nonlinear dynamo theory,'' submitted to Phys. Rep., 
astro-ph/0405052, (2004) 

\bibitem{bbs02}
A. Brandenburg, A. Bigazzi, K. Subramanian, 
Astron. Nachr. {\bf 323} 99 (2002)

\bibitem{taylor86}
J.B. Taylor, Rev. Mod. Phys., {\bf 58}, 741 (1986) 

\bibitem{gp97}
L. Golub \& J. Pasachoff {\it Solar Corona},
(Cambridge: Cambridge Univ. Press, 1997)

\bibitem{withbroe77} 
G.L. Withbroe,  \& R.W. Noyes, Ann.\ Rev.\ Astron.\
Astrophys.\ {\bf 15}, 363, (1977)

\bibitem{sz}
C.J. Schrijver  \& C. Zwaan, {\it Solar and Stellar Magnetic Activity},
(Cambridge: Cambridge Univ. Press, 2000)

\bibitem{wang03}  Y.-M. Wang \& 
N.R. Sheeley, Astrophys. J., {\bf 599}, 1404 (2003)

\bibitem{galeev79} 
A.A. Galeev, R. Rosner, G.S. Vaiana, Astrophys. J., {\bf 229} 318 (1979)

\bibitem{fr93} 
G.B. Field \& R.D. Rogers, R.D. Astrophys. J., {\bf 403} 94 (1993) 

\bibitem{haardt93}F. Haardt  \& L. Maraschi, Astrophys. J., {\bf 413}, 507, (1993)

\bibitem{wang04}
J. Wang, K. Watarai,  \& S. Mineshige, Astrophys. J. Lett., {\bf 607}, L107 (2004)

\bibitem{fm99} E.D. Feigelson, 
\& T. Montmerle, Ann. Rev. Astron. Astrophys., {\bf 37}, 363 (1999)

\bibitem{lpp}
S.H.  Lubow,  J.C.B. Papaloizou, J.E. Pringle, Mon. Not. R. Astron. Soc., 267, 235 (1994).

\bibitem{blackmantan04} E.G. Blackman \& J.C. Tan,
Astrophys. \& Space Sci., in press (astro-ph/0306424)

\bibitem{bp82}R.D. Blandford \& 
D.G. Payne, Mon. Not. R. Astron. Soc., {\bf 199}, 883 (1982)

\bibitem{love87} R.V.E. Lovelace, J.C.L, Wang,  \& M.E. 
Sulkanen, M.E. Astrophys. J., {\bf 315}, 504 (1987)

\bibitem{meier01} D.L. Meier,  
S. Koide,  \& Y. Uchida, \  Science, {\bf 291}, 84  (2001)

\bibitem{ouyed03}R. Ouyed,  
D.A. Clarke, \& R.E. Pudritz, Astrophys. J., {\bf 582} {292} (2003) 

\bibitem{koide04} S. Koide,  Astrophys. J. Lett., {\bf 606}, L45 (2004)

\bibitem{campbell99}
C.G. Campbell, Mon. Not. R. Ast. Soc., {\bf 306}, 307 (2000)

\bibitem{vrb04} B. Von Rekovski \& A. Brandenburg, Astron. \& Astrophys.,  
{\bf 420}, 17 (2004)

\bibitem{lovelace02} R.V.E. Lovelace, H. Li, A.V. Koldoba, 
G.V. Ustyugova, \& M.M. Romanova, M.~M. Astrophys. J. Lett., 
{\bf 572}, 445 (2002)

\bibitem{ms00} K.A. Miller and J.M. Stone, Astrophys. J. Lett., {\bf 534}, 398 (2000)

\bibitem{pudritz81}R.E. Pudritz, Mon. Not. R. Astron. Soc., 
{\bf 195}, 897 (1981)

\bibitem{rrs97} M. Reyes-Ruiz, 
\& T.F. Stepinski, T.~F.\  Mon. Not. R. Astron. Soc., {\bf 285}, 501 (1997)

\bibitem{rrs99} M. Reyes-Ruiz, 
\& T.F. Stepinski, Astron. \& Astrophys., {\bf 342}, 892 (1999)


\bibitem
{yousef03}
T.A. Yousef \& A. Brandenburg, Astron. \& Astrophys., {\bf 407}, 7 (2003).


\bibitem{woltjer}
L. Woltjer, Proc. Nat. Acad. Sci., {\bf 44}, 489 (1958)

\bibitem{berger84}
M.A. Berger \&  G.B. Field,  J. Fluid Mech., {\bf 147}, 133 
(1984)

\bibitem{vc01} E.T. Vishniac, \& 
J. Cho, Astrophys. J {\bf 550}, 752  (2001)

\bibitem{bs04} A. Brandenburg \& C. Sandin, 
``Catastrophic alpha quenching alleviated by helicity flux and shear,''
submitted to Astron. \& Astrophys. (2004)



\eject

\noindent{{\bf Figure 1}:
Driven bihelical magnetic relaxation for $c_1=0$. 
Here $h_2=-h_3 =1$, for all time, 
$k_1=1$, $k_2=5$, $k_3=20$ and $h_1(t=0)=0.01$.
(a) $h_1$,
where $R_M = R_V =1000$ (dashed) $R_M = R_V= 6000$ (solid).
The saturation value of $h_1$ depends linearly on $R_M$.
(b) Same as (a) but for early times. The kinematic 
regime ends where the two curves split. Note that
just after the kinematic regime the $R_M=6000$ 
curve is below that of $R_M=1000$ in contrast to the late-time regime in (a).
(c) $h_2$ (short dashed, straight line) and $h_2^V$ (long dashed).
The curves for $R_M=6000$ and $R_M=1000$ indistinguishably overlap.
(d) $h_3$ (straight dashed line) and $h_3^V$ (oscillating solid line). Again negligible
variation for $R_M=1000$ and $R_M=6000$ cases.
(e) Dimensionless EMFs:
${\overline Q}$ (solid humped curve), ${\overline q}$ (long dashed), 
and ${q}_2$ (short dashed) for $R_M=1000$.
(f) $h_1$ when   $h_3^V << 1$ is artificially 
enforced for $R_M=1000$, 
and multiplying the $q_2$ term in (\ref{Q2})
by 1 (dashed) and 0.5  (solid) to show the relative effect
on $h_1$.

\noindent{{\bf Figure 2}:  Same as Fig. (1a) but with
different values of the $h_1$ loss coefficient 
$c_1$, and $h_1(0)$. 
(a) $c_1=0.1$, $h_1(0)=0.01$
(b) $c_1=1$}, $h_1(0)=0.01$.
These late time curves  are insensitive to the choice of $h_1(0)$.

\noindent{{\bf Figure 3}:  Kinetic helicity
driven dynamo. Here the system is driven such that  $h_2^V=1$
for all times, and a seed $h_1(0)=0.01$. 
These curves show that Eqs. (\ref{h1}-\ref{h3v})
reproduce the results of Ref. \cite{bb02}
for the appropriate initial conditions.
(a) Late time solution of $h_1$ for $R_M=R_V=1000$ (dashed)
and $R_M=R_V=6000$ (solid). 
(b) early time solution for $h_1$.

\noindent{{\bf Figure 4}:  
Unihelical magnetic  relaxation.
Here the system is driven with $h_2=1$ and $h_3=0$
for all times, with a seed $h_1(0)=0.01$.
These curves show that the generalized set
of equations reproduces unihelical relaxation of of Ref. \cite{bf04}.
(a) $h_1$ for $R_M=R_V=1000$ (dashed)
and $R_M=R_V=6000$ (solid). (b) same as (a) but at early times
(c) $h_1$ for $R_M=R_V=1000$ (solid)
and $R_M=10R_V=1000$ (dashed) (d) same as (c) for early times.

\noindent{{\bf Figure 5}:
Free bihelical magnetic relaxation. 
Here $h_1(0)=0.01$, $h_2(0)=-h_3(0)=1$ but unlike the case of Fig. 1,
there is no subsequent driving.
(a) $h_1$,
where $R_M = R_V =2000$ (dotted) $R_M = R_V =10000$ (dashed) 
and $R_M = 10000$ and $R_V=2000$ (solid).
Note the slower decay for larger $R_M$ and $R_V$. 
(b) Same as (a) for early times. Note  the insensitivity
to $R_V$ at these times since the dotted and solid curves
defined in (a) overlap.
(c) 
$h_2$ for $R_M=R_V=2000$ (thin solid);
$h_2^V$ for $R_M=R_V=2000$ (short-dashed);
$h_2$ for $R_M=R_V=10000$ (thick solid);
$h_2^V$ for $R_M=R_V=10000$ (long-dashed).
Note that $h_2$  initially decays  and $h_2^V$ rises
and then the curves decay together. The solutions for 
$h_2$ and $h_2^V$ are insensitive
to $R_V$ for a given $R_M$. Note that $h_2,\ h_2^V$ 
actually deplete faster for larger $R_M$ (see (e) and text).
(d) $h_3$ and $h_3^V$. At early times (barely
visible), $|h_3|$ decays and $|h_3^V|$ increases and then
the curves nearly overlap in decay:
$h_3$ and $h_3^V$ for $R_M=R_V=2000$ (top);
$h_3$ and $h_3^V$ for $R_M=R_V=10000$ (bottom)
$h_3$ and $h_3^V$ for $R_M=10000$, $R_V=2000$ (middle).
(e) Blow-up of $(d)$ to show the offset between $h_3^V$ and $h_3$
with the $h_3^V$ curves are the solid curves 
on the right of each pair, and $h_3$ is the left dashed
curve in each pair. The pairs from left
to right are $h_3$ and $h_3^V$ for $R_M=R_V=2000$;
$h_3$ and $h_3^V$ for $R_M=R_V=10000$;
$h_3$ and $h_3^V$ for $R_M=10000$, $R_V=2000$. 
The offset accounts for the weak $h_3^V$ driven dynamo
that grows $h_1$ slightly beyond its $h_1=1/2$ kinematic growth value.
(f) Dimensionless EMFs: 
${\overline q}$ for $R_M=R_V=10000$ (thick solid), 
${\overline q}$ for $R_M=R_V=2000$ (long dashed), 
${q}_2$ for $R_M=R_V=10000$ (thin solid), 
${q}_2$ for $R_M=R_V=2000$ (short dashed). 

\eject
\vspace{0cm} 
{\hspace{-4cm}
\vspace{-2cm}  
\epsfxsize11.6cm
\epsffile{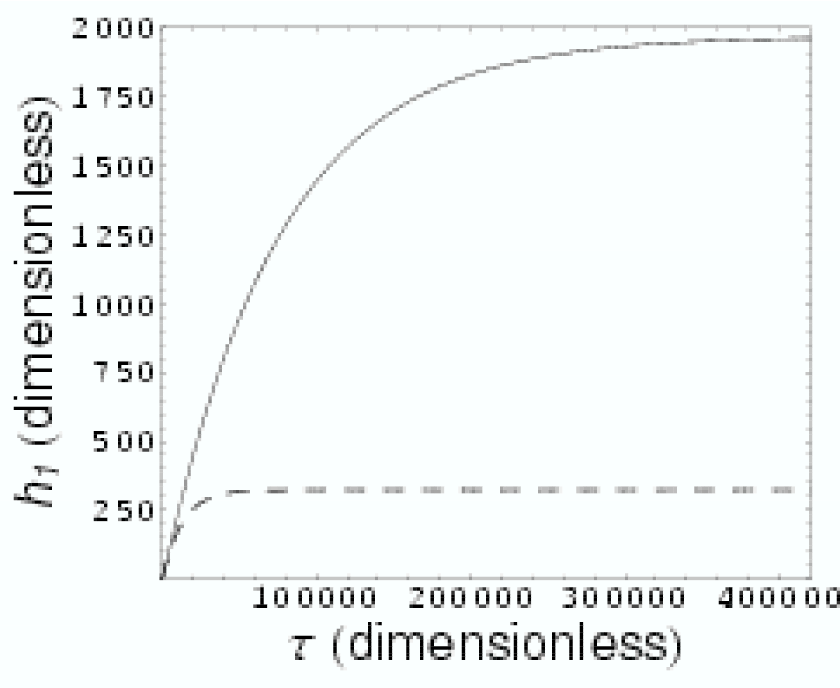} \hspace{-4cm} 
\epsfxsize11.6cm \epsffile{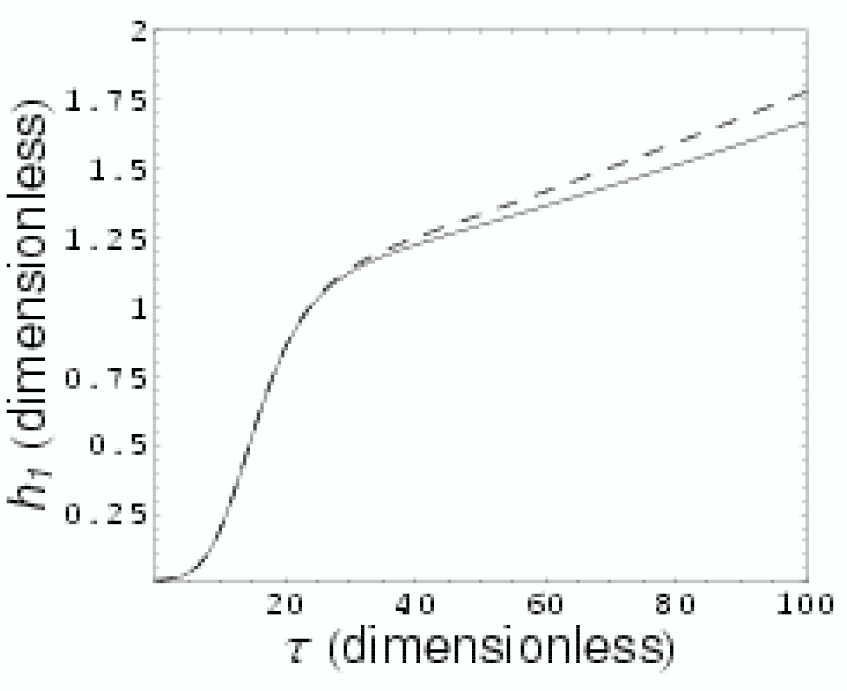}}
\vspace{-2cm} 
{\hspace{-4cm} \epsfxsize11.6cm \epsffile{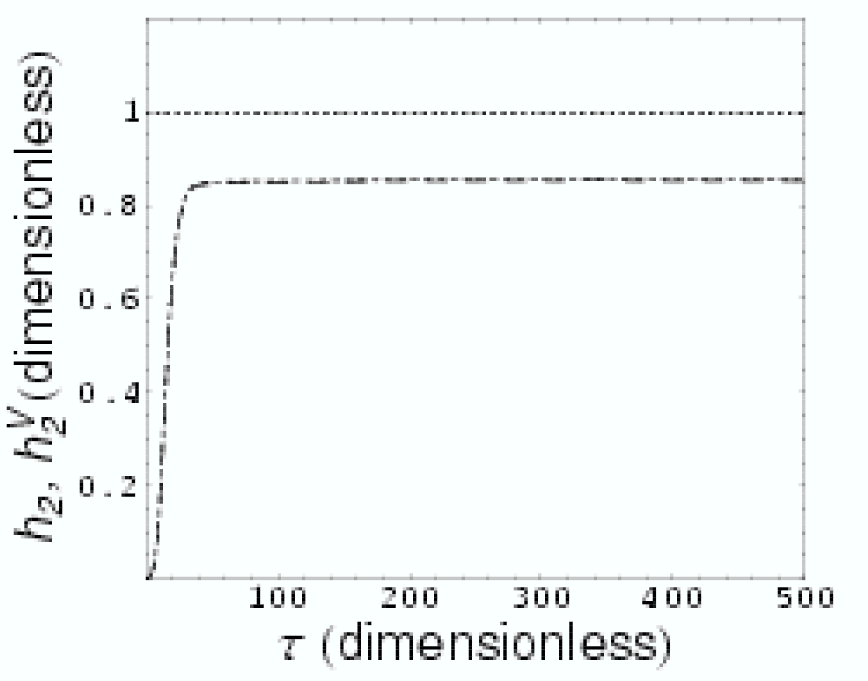} \hspace{-4cm} 
\epsfxsize11.6cm  \epsffile{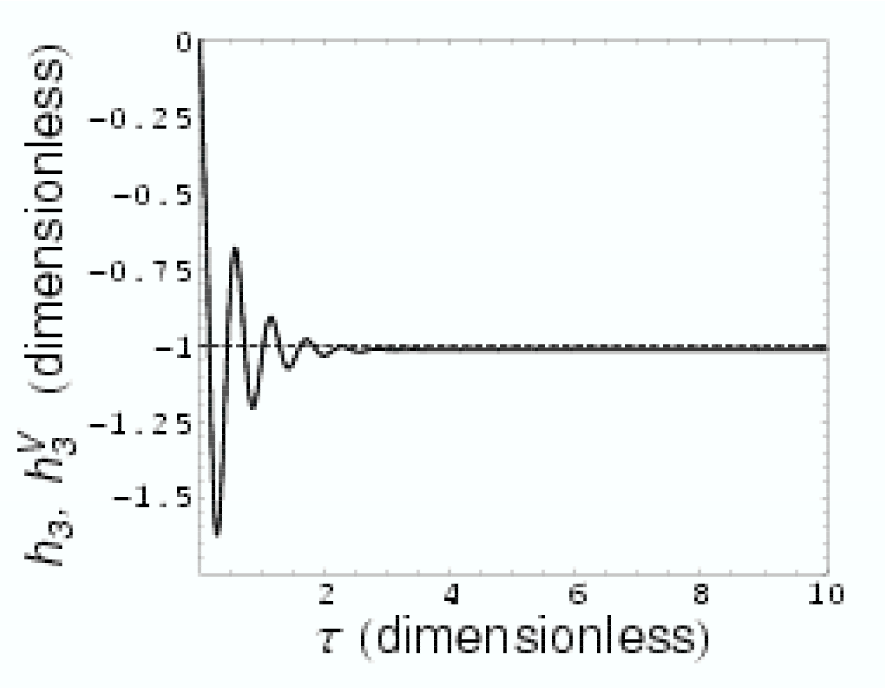}}
\vspace{-2cm} 
{\hspace{-4cm} \epsfxsize11.6cm \epsffile{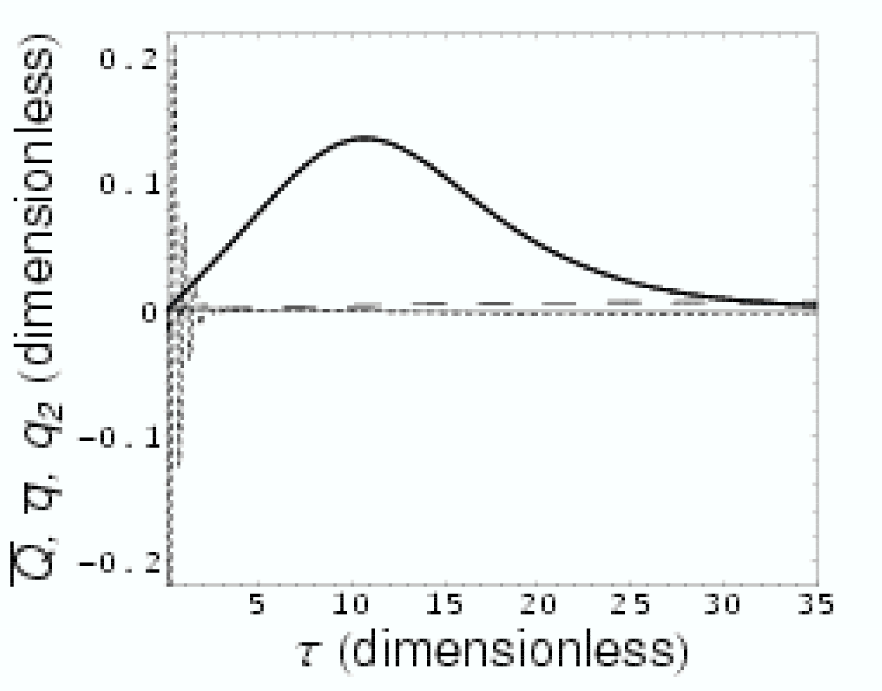}\hspace{-4cm} 
{\epsfxsize11.6cm \epsffile{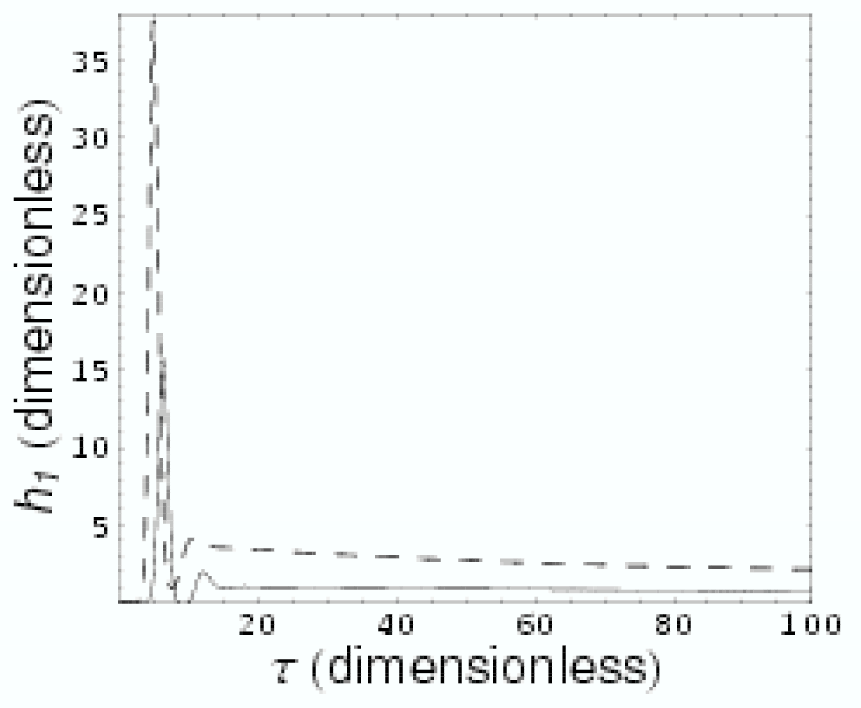}}

{Figure 1}
\eject

\vspace{1cm} \hbox to \hsize{\epsfxsize11.6cm
\epsffile{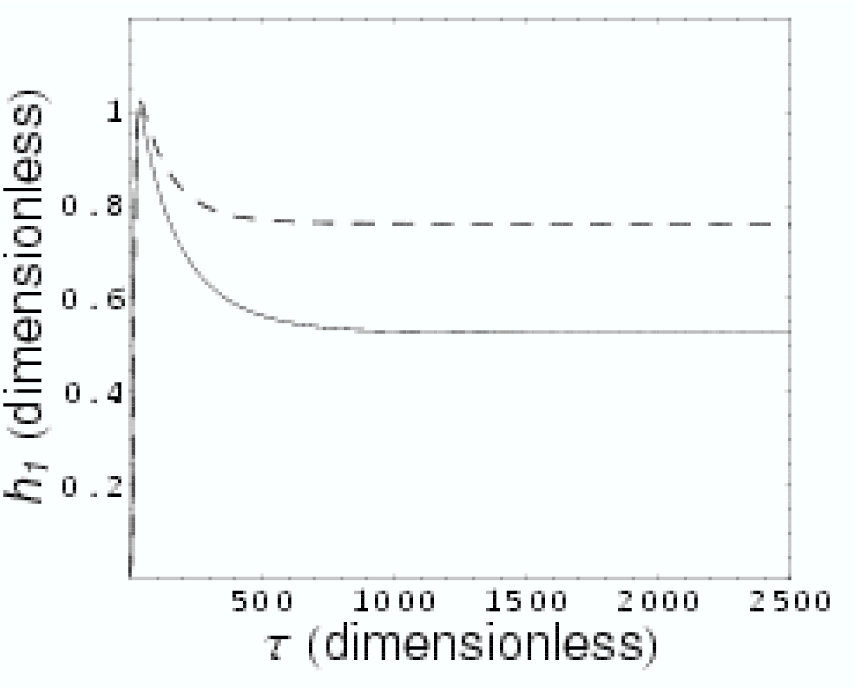} \hspace{-4cm} \epsfxsize11.6cm \epsffile{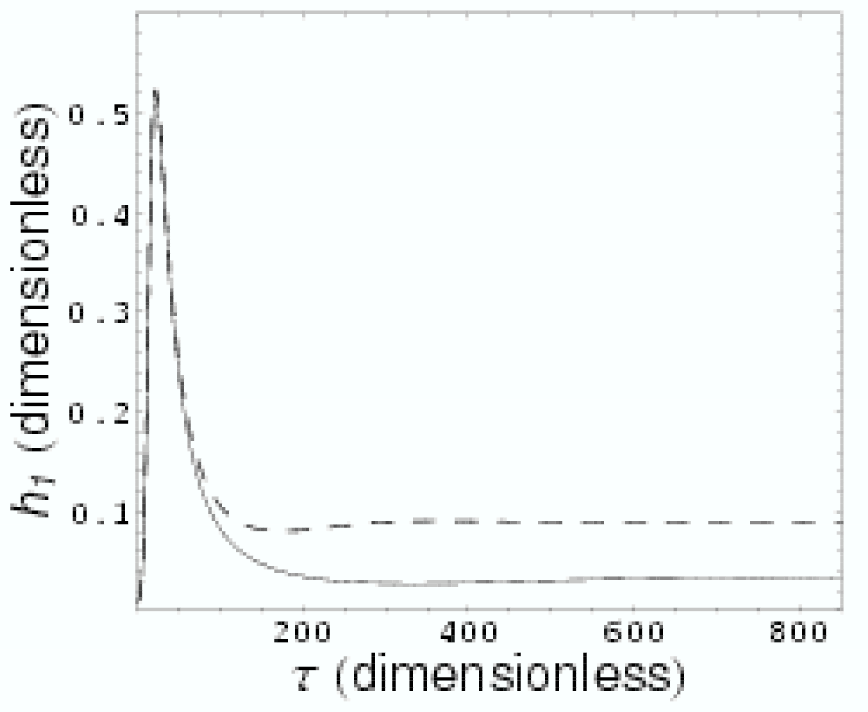}}
\vspace{0cm} \hbox to \hsize{\epsfxsize10.5cm
}

{Figure 2}

\eject

\vspace{-.1cm} \hbox to \hsize{\epsfxsize11.6cm
\epsffile{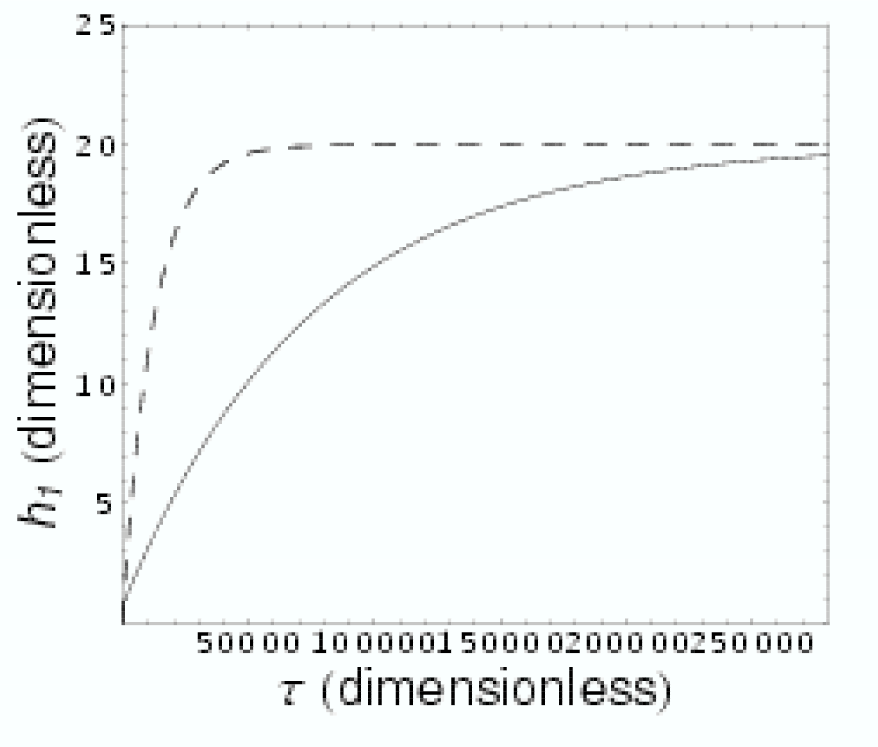} \hspace{-2cm}\epsfxsize11.6cm \epsffile{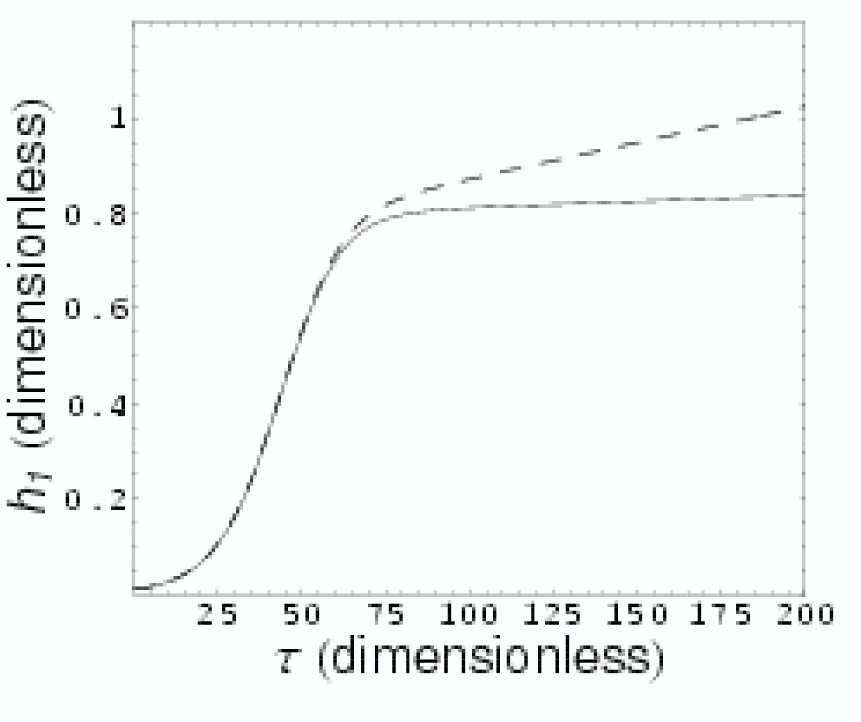}}

{Figure 3}

\eject
\vspace{-.1cm} \hbox to \hsize{\epsfxsize11.6cm
\epsffile{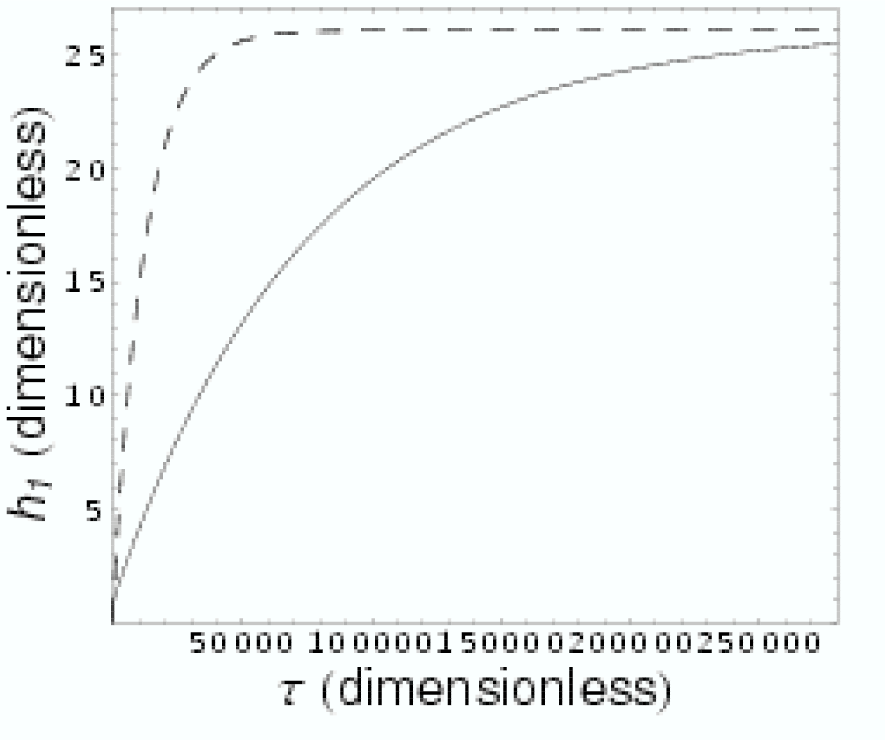} \hspace{-4cm}
\epsfxsize11.6cm \epsffile{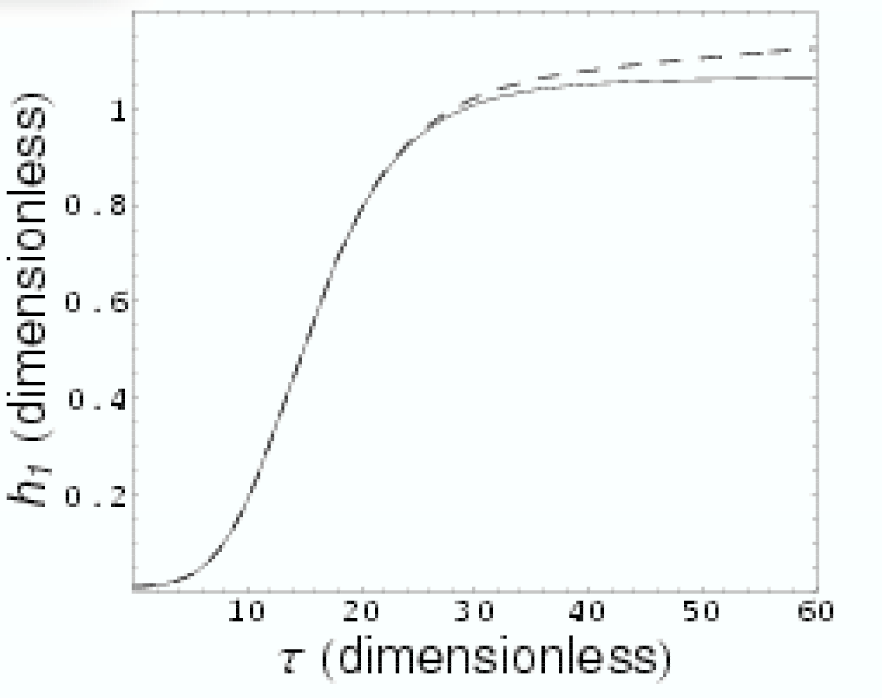}}
\vspace{-.1cm} \hbox to \hsize{\epsfxsize11.6cm
\epsfxsize11.6cm \epsffile{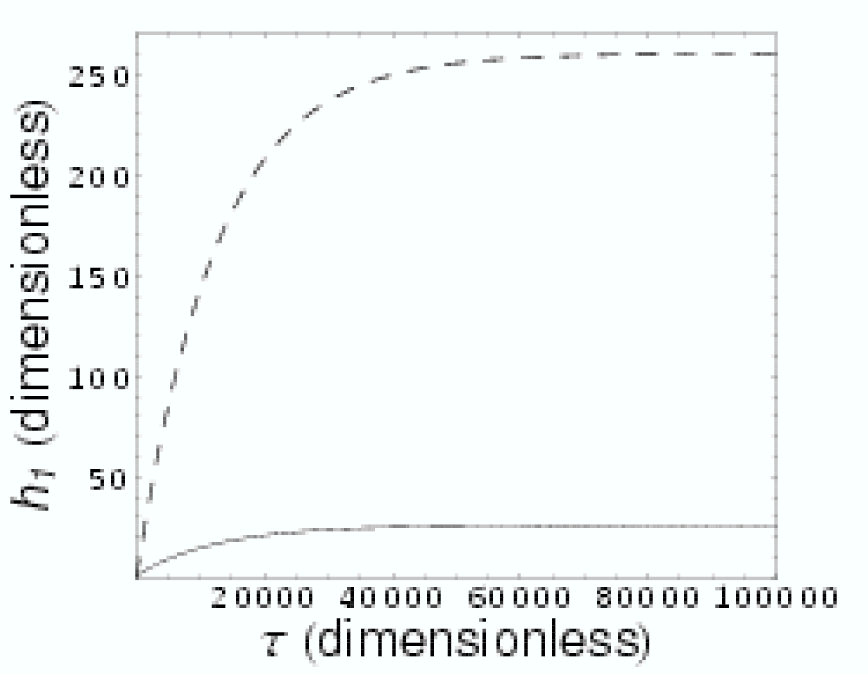} 
\hspace{-4cm}\epsfxsize11.6cm \epsffile{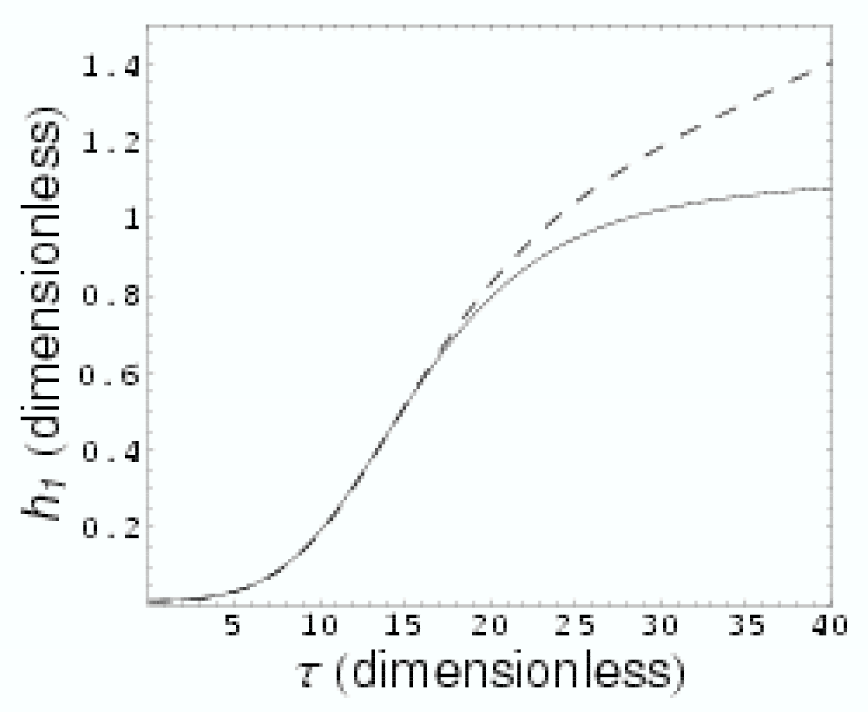}}

{Figure 4}

\eject
\vspace{0cm} 
{\hspace{-4cm} \epsfxsize11.6cm \epsffile{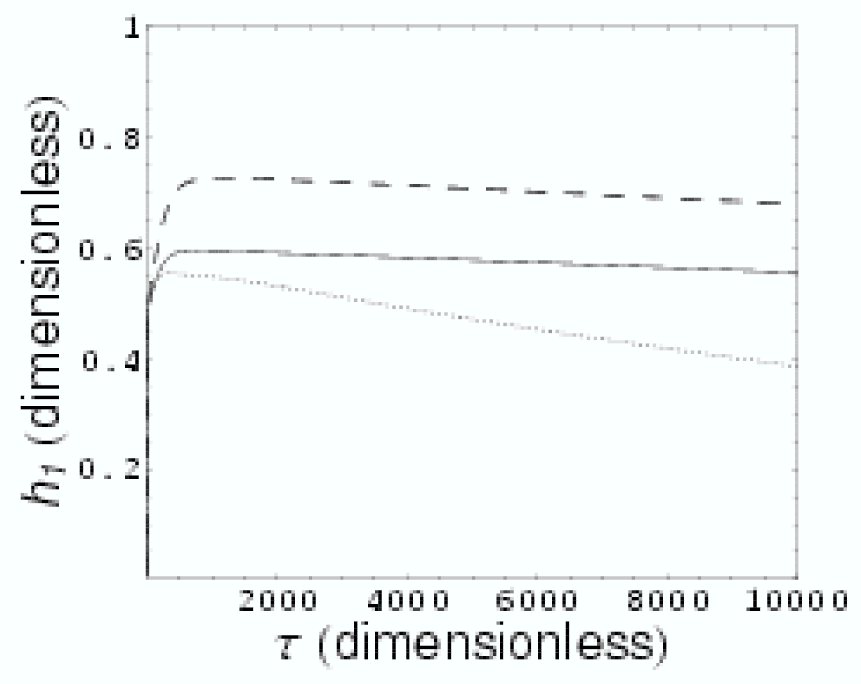}
\hspace{-4cm} \epsfxsize11.6cm \epsffile{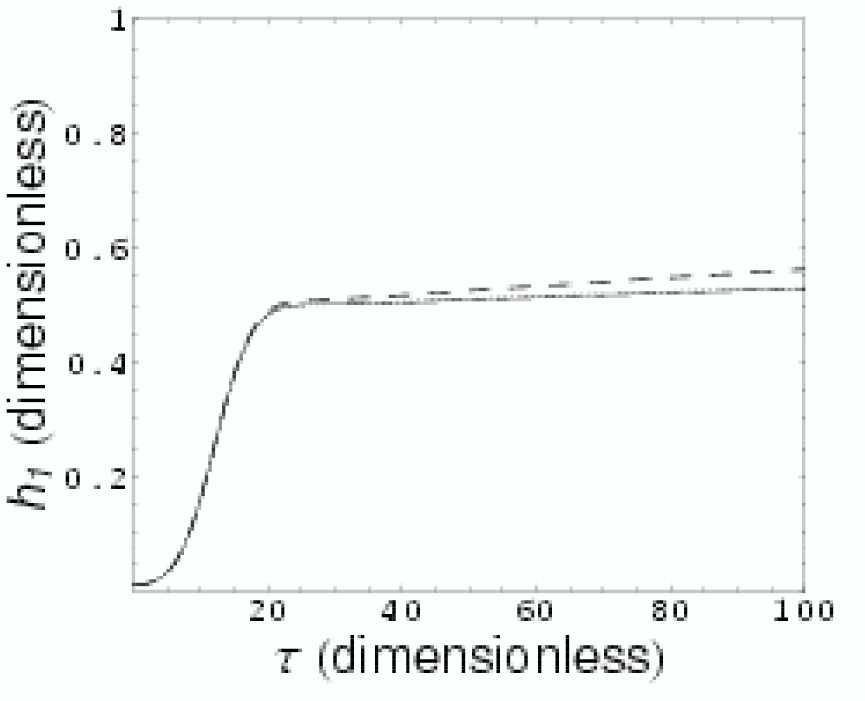}}
\vspace{-2cm} 
{\hspace{-4cm} \epsfxsize11.6cm \epsffile{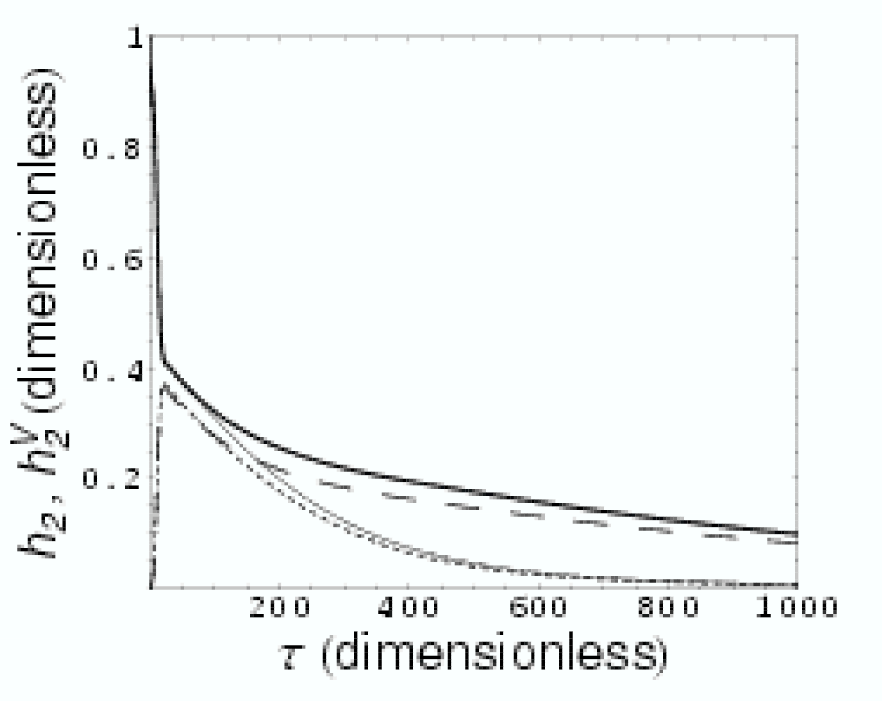} 
\hspace{-4cm} \epsfxsize11.6cm \epsffile{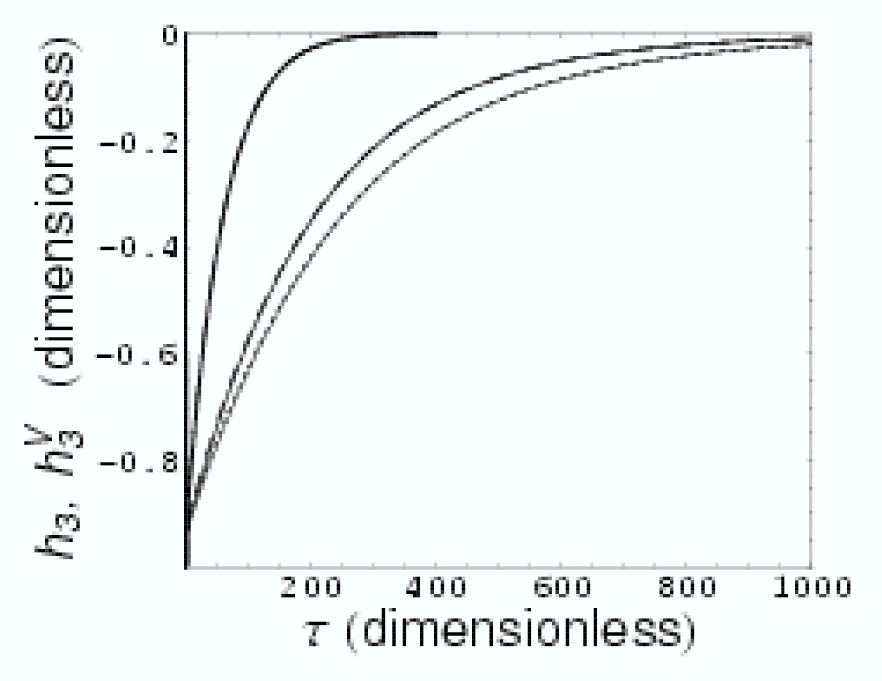}}
\vspace{-2cm} 
\hspace{-4cm} 
\epsfxsize11.6cm  \epsffile{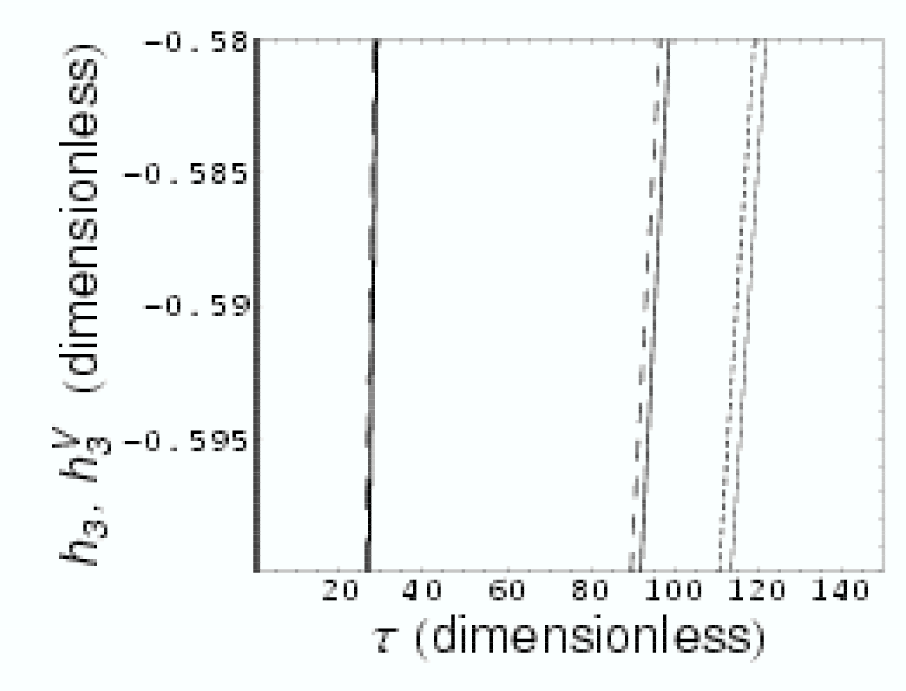}
{\hspace{-4cm}
\epsfxsize11.6cm  \epsffile{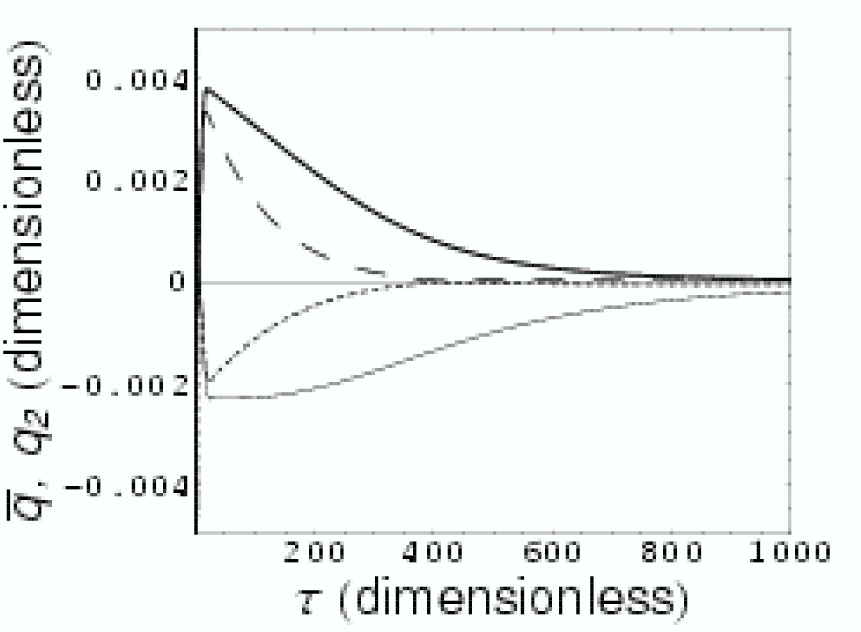}}

\vspace{1cm}

{Figure 5}

\eject

\end{document}